\documentclass[useAMS,usenatbib]{mn2e}
\usepackage{epsfig,color}
\usepackage{bm,graphicx,morefloats}
\usepackage{amsmath}

\title[Pop~III initial mass function]
      {Building up the Population~III initial mass function from cosmological initial conditions}

\author[A. Stacy et al.]
       {Athena Stacy$^{1}$\thanks{E-mail: athena.stacy@berkeley.edu}, Volker Bromm$^{2}$, and Aaron T. Lee$^{1}$\\
        $^{1}$University of California, Berkeley, CA 94720, USA \\
        $^{2}$Department of Astronomy, University of Texas, Austin, TX 78712, USA
         }

\pagerange{\pageref{firstpage}--\pageref{lastpage}}
\pubyear{2014}

\begin{document}

\maketitle
\topmargin-1cm

\label{firstpage}

\begin{abstract}
We simulate 
the growth of a Population III stellar system, starting from cosmological initial conditions at $z=100$.  
We
follow the formation of a minihalo and the subsequent collapse of its central gas to high densities, resolving scales as small as $\sim$\,1 AU.
Using sink particles to represent the growing protostars, we model the growth of the photodissociating and ionizing region around the first sink, 
continuing the simulation for $\sim$\,5000 yr after initial protostar formation.
Along with
the first-forming sink, several tens of secondary sinks form before an ionization front develops around the most massive star.  
The resulting cluster has high rates of sink formation, ejections from the stellar disc, and sink mergers during the first $\sim$\,2000 yr, before the onset of radiative feedback.  
By this time 
a warm $\sim 5000$~K phase of neutral gas has expanded to roughly the disc radius of 2000 AU, slowing mass flow onto the disc and sinks. 
By 5000 yr
the most massive star grows to 20~M$_{\odot}$, while the total stellar mass approaches 75~M$_{\odot}$.  
Out of the $\sim$\,40 sinks, approximately 30 are low-mass ($M_*< 1$ M$_{\odot}$),
and if the simulation had resolved smaller scales an even greater number of sinks might have formed.
Thus, protostellar radiative feedback is insufficient to prevent rapid disc fragmentation and the formation of a high-member Pop III cluster before an ionization front emerges.  
Throughout the simulation, the majority of stellar mass is contained within the most massive stars, further implying that the Pop III initial mass function is top-heavy.
\end{abstract}

\begin{keywords}
stars: formation - Population III - galaxies: formation - cosmology: theory - first stars - early Universe 
\end{keywords}

\section{Introduction}

After recombination, when the photons of the cosmic microwave background (CMB) were released,
the Universe exhibited a very uniform structure and contained no point sources of light. As gravitational perturbations grew, the first stars and galaxies formed and became early drivers of cosmic evolution, towards the highly structured state we observe today.  
Prior to the formation of the first stars, the so-called Population~III (Pop~III),
no metals or dust existed to facilitate the cooling and condensation of gas into stars 
as in contemporary star formation. 
Primordial star formation was instead driven by cooling through H$_2$ line emission. Pop III stars are believed to have initially formed at $z\sim20-30$ in small dark matter (DM) haloes of mass $\sim10^6$ M$_{\odot}$. These `minihaloes' were the first structures whose constituent gas had a sufficient abundance of H$_2$, such that the energy release due to rovibrational transitions enabled continued collapse (e.g. \citealt{haimanetal1996,tegmarketal1997,yahs2003}). A gravitationally-bound gas core of 
$\sim$\,1000 M$_{\odot}$ developed at the centre of each star-forming minihalo, and this core provided the reservoir from which primordial protostars formed and grew.

The radiation subsequently emitted during the lifetime of the first stars, and the metals they released through supernova (SN) explosions and stellar winds, left a crucial imprint on their environment.  
Sufficiently massive Pop III stars carved out the first H{\sc ii} regions (e.g. \citealt{kitayamaetal2004,syahs2004,whalenetal2004,alvarezetal2006,johnsongreif&bromm2007}).
Some of these same massive Pop III stars also left behind the first metal-enriched SN remnants 
(e.g.
\citealt{moriferrara&madau2002,brommyoshida&hernquist2003,wada&venkatesan2003,normanetal2004,tfs2007,greifetal2007,greifetal2010,wise&abel2012,maioetal2011}; recently reviewed in \citealt{karlssonetal2013}).
In the wake of Pop III stars, the first galaxies emerged, forming within higher-mass $\sim 10^{7-8}$ M$_{\odot}$ `atomic cooling' haloes to continue the process of reionizing the Universe and enriching the intergalactic medium (IGM).  

As in the metal-enriched case, Pop III stars undergo a great variety of stellar deaths.
Specifically, Pop~III stars in the mass range of 140--260M$_{\odot}$ die as pair-instability supernovae (PISNe).
A PISN event, which disrupts the entire star and leaves behind no compact remnant (e.g. \citealt{hegeretal2003}), gives rise to violent
mechanical feedback and ejects 
 $\sim$\,100 M$_{\odot}$ 
of heavy elements into the surroundings.
 In contrast, 
a lower-mass progenitor will undergo a less-energetic core-collapse SN (CCSN).  
Pop III stars with main-sequence masses in the ranges of 40-140 M$_{\odot}$ or $>260$ M$_{\odot}$ collapse directly into black holes (BHs).  
Such BH remnants will disperse no, or only few, metals, but they will continue to irradiate the Universe in X-rays as they accrete mass
(e.g. \citealt{jeonetal2012, hummeletal2015}). In addition, 
  some of these BH remnants may have powered collapsar gamma-ray bursts if their Pop III progenitor had sufficient angular momentum.
(\citealt{woosley1993,stacyetal2011,stacyetal2013}). 
 Finally, 
some Pop~III stars may be low-mass ($< 0.8$ M$_{\odot}$) and long-lived such that they survive to the present day, thus rendering them, in principle, observable within the Milky Way or nearby dwarf galaxies. Indeed, if ongoing surveys such as {\it Gaia} or with the {\it SkyMapper} telescope failed to find any metal-free stars, increasingly strong constraints could be placed on the lower mass of Pop~III star formation (e.g. \nocite{hartwigetal2015a} Hartwig et al. 2015a).  

The rate at which these various stellar end stages occurred, and the subsequent effect of Pop III stars on later generations of star formation, depends upon the Pop III initial mass function (IMF).
Pop III stars at typical redshifts of $z\ga 20$ and non-extreme masses are too faint to be individually detectable by even next-generation telescopes such as the {\it James Webb Space Telescope (JWST}; \citealt{gardneretal2006}).  
We instead employ numerical simulations to constrain the Pop III IMF. 
 Early simulations indicated that the IMF is dominated by stars of very high mass
($\ga 100$ M$_{\odot}$; e.g. \citealt{abeletal2002,brommetal2002,bromm&loeb2004,yoh2008}). 
More recent simulations have found that fragmentation of the central gas configuration will instead allow for a wider range of Pop~III masses
 (e.g. \nocite{clarketal2008,clarketal2011a} Clark et al 2008; 2011a; \citealt{turketal2009,stacyetal2010,stacy&bromm2013, stacy&bromm2014}).  
 The mass functions found in these simulations were generally logarithmically flat and top-heavy. 
 This is in strong contrast to the bottom-heavy IMF observed in the Milky Way and Local Group dwarf galaxies (e.g. \citealt{salpeter1955, chabrier2003}), where the majority of stellar mass is found within low-mass stars ($\la$ 1 M$_{\odot}$). 
  One key reason for this difference is
 the distinct chemo-thermal processes which occur in primordial gas, leading to less rapid cooling than in enriched gas which can release thermal energy through dust and metal-line emission. The warmer primordial gas has a higher Jeans mass $M_{\rm J}$, and this ultimately leads to more rapid accretion rates than that found in enriched gas.  
While these simulations did not include effects of protostellar feedback, other work has found similar fragmentation rates and mass functions even when the gas is heated through radiation (Clark et al. 2011b, \nocite{clarketal2011b} \citealt{greifetal2011,smithetal2011}).  
 
Existing simulations thus presented preliminary mass functions, but the final Pop~III masses may be reached when ionizing radiation reverses mass infall onto the stars and halts accretion (e.g. \citealt{mckee&tan2008}).
\cite{stacyetal2012} explored this process through a three-dimensional cosmological simulation which followed the growth of a protostellar system under photodissociating and ionizing feedback.  
This simulation, which resolved scales down to $\sim$\,50 AU, resulted in the formation of a Pop~III binary whose members had masses $\sim$\,20 and 10 M$_{\odot}$ after 5000~yr of accretion. By this time the accretion rate had declined by an order of magnitude, and extrapolating to a Kelvin-Helmholtz (KH) time of $\sim$\,10$^5$ yr implied the largest protostar would have grown to 30 M$_{\odot}$ by the time it reached the main sequence.  However, this study could not resolve fragmentation on $<50$ AU scales.
A two-dimensional simulation of comparable spatial resolution by \cite{hosokawaetal2011} followed the growth of a Pop~III star to 40 M$_{\odot}$, after which radiative feedback shuts off any further accretion. Further work by \cite{hiranoetal2014} performed similar two-dimensional studies for approximately 100 different host minihaloes, finding a much larger Pop~III mass range of 10-1000 M$_{\odot}$.   However, disc fragmentation and the three-dimensional effects of non-axisymmetry could not be followed in these calculations.  
\cite{susaetal2014} used three-dimensional simulations and a minimum resolution radius of 30 AU to study Pop III star formation within several tens of haloes under photodissociating feedback, finding a wide range of stellar masses and multiplicities, with a mass spectrum ranging from 1 to $>$ 100 M$_{\odot}$.
\cite{hosokawaetal2016} also numerically studied several different minihalo environments using a resolution of $\sim$\,30 AU, but included both photodissociating and ionizing feedback. They similarly found a wide range of masses, from $\sim$\,10 to several hundred M$_{\odot}$ across different minihaloes, but no binary or multiple formation.  

In this paper we present a three-dimensional numerical simulation to follow the growth of a Pop III stellar system from cosmological initial conditions, while also resolving nearly protostellar scales ($\sim$\,100 R$_{\odot}$) and accounting for the effects of dissociating and ionizing radiation.  
We additionally account for the effects of radiation pressure from ionizing photons.
Recent work has found that 100 R$_{\odot}$ ($\sim$\,1 AU) is approximately the maximum radius reached by a Pop III protostar during its pre-main sequence evolution (see e.g. \citealt{hosokawaetal2010, smithetal2012}).  Resolving these small scales corresponds to resolving a maximum density of 10$^{16}$ cm$^{-3}$, at which point we continue the gas evolution for a further 5000~yr by employing sink particles to represent the growing protostars.
At this high density the gas is quickly 
 becoming optically thick
to continuum radiation, which occurs at 10$^{18}$ cm$^{-3}$, such that the gas is unlikely to undergo further fragmentation on sub-sink scales (\citealt{yoh2008}).  
This enables us to resolve the protostar formation during the initial collapse phase.

While the numerical methodology is similar to that used in \cite{stacy&bromm2014}, the protostars studied in that work did not grow large enough to develop an I-front. Here, we specifically choose a minihalo which is known, from previous lower-resolution simulations, to host a rapidly growing and massive protostellar system whose largest protostar will grow sufficiently massive to develop an HII region. We follow the evolution until radiative feedback has effectively shut off accretion onto the protostellar disc. The masses reached by the sinks at the end of this simulation can therefore be considered close to the final mass they will reach in their lifetime, 
 as the stellar cluster's accretion rate has
effectively fallen to zero.
This computation thus models the fragmentation of primordial gas and the growth of Pop III protostars
with improved physical realism. We are thus getting closer to the asymptotic goal of building up the Pop~III IMF from ab initio cosmological initial conditions.  
The structure of our paper is as follows.
In Section 2, we describe our numerical methodology, while we present our protostellar evolution model in Section~3. 
The early gas evolution is discussed in Section 4, while the onset of radiative feedback is discussed in Section 5.  We describe the resulting Pop III cluster characteristics in Section 6, and we summarize and conclude in Section 7.

\section{Numerical Methodology}

\subsection{Initial Setup and Refinement Technique}

This current work employs the same cosmological simulation as in \cite{stacy&bromm2013}, 
and the setup and refinement techniques were described in this previous paper. 
For completeness and convenience of the reader, we briefly summarize the key aspects here.  
We employed {\sc gadget-2} (\citealt{springel2005}) and used a 1.4 Mpc (comoving) box containing 512$^3$ SPH gas particles and 512$^3$ DM particles, initialized at $z=100$.  
The initialization is performed according to a $\Lambda$CDM cosmology with $\Omega_{\Lambda}=0.7$, $\Omega_{\rm M}=0.3$, $\Omega_{\rm B}=0.04$, $\sigma_8=0.9$, and $h=0.7$.
In this un-refined simulation, gas particles each had mass $m_{\rm sph} = 120$ M$_{\odot}$, and DM particles each had mass  $m_{\rm DM} = 770$ M$_{\odot}$. 

The cosmological simulation was run until the first ten minihaloes were formed between $z\sim30$ and $z\sim20$.  We next performed ten re-simulations, again initialized at $z=100$.  Each of these ten re-simulations had a cubical region of increased refinement centred on the future formation site of one of the ten minihaloes found in the original simulation.  The highly refined region contained SPH and DM particles with $m_{\rm sph} = $1.85~M$_{\odot}$ and $m_{\rm DM} = 12$~M$_{\odot}$.

Every re-simulation was run until the gas in its refined minihalo reached a density of $5 \times 10^7$ cm$^{-3}$.  
At this point, we cut out all particles outside of 10 (physical) pc from the densest particle.  
This allowed the computational time and memory to be focused only on the central star-forming gas.    
Before continuing the cut-out simulation, we used a particle-splitting technique in which the mass of each parent SPH particle was sub-divided into 256 child particles, such that each child particle has mass $m_{\rm sph}=7.2\times10^{-3}$ M$_{\odot}$.  Each child particle receives the chemical abundances, velocity, and internal energy of its parent particle.  This SPH splitting technique has been previously used by, e.g., \cite{bromm&loeb2003} and \nocite{clarketal2011b} Clark et al. 2011b.  We refer the reader to \cite{chiaki&yoshida2015} for more recently developed splitting techniques,
 and we refer the reader to \cite{stacy&bromm2013} for resolution tests of the technique used for this work.  
 
For this work, we focus on the most rapidly accreting minihalo of \cite{stacy&bromm2013}.  
In our new simulation, when the gas within this minihalo has attained a maximum density of 10$^8$ cm$^{-3}$, we
further split each particle into 12 child particles.
We reduce the cut-out box to a smaller size of 3 pc across, again centred on the densest gas particle.    
In this final and most highly-resolved simulation, each SPH particle thus has mass $m_{\rm sph}=6\times10^{-4}$ M$_{\odot}$, yielding a mass resolution of $M_{\rm res}\simeq 1.5 N_{\rm neigh} m_{\rm SPH} \simeq 3 \times 10^{-2} $M$_{\odot}$, where $N_{\rm neigh}\simeq 40$ is the typical number of particles in the SPH smoothing kernel (e.g. \citealt{bate&burkert1997}).  
This allows us to follow the gas collapse to densities of 10$^{16}$ cm$^{-3}$.

\subsection{Chemistry, Heating, and Cooling}

The chemistry and thermal networks have been described in previous work (e.g. \citealt{greifetal2009, stacyetal2012}).   
For completeness we summarize the key points here.

Each SPH particle is initialized with the temperature and a set of chemical abundances appropriate at $z=100$.  More specifically, the chemical species are H, H$^{+}$, H$^{-}$, H$_{2}$, H$_{2}^{+}$, He, He$^{+}$, He$^{++}$, e$^{-}$, and the three deuterium species D, D$^{+}$, and HD.  Each particle also has a variable adiabatic index $\gamma$ which is updated according to the evolving temperature and abundances.  The value of 
$\gamma$ ranges between 
 $\sim$\,1.3 - 1.4 
for molecular gas and $5/3$ for atomic and ionized gas.

At each timestep, after gravitational and hydrodynamic forces are updated, abundances and temperatures are evolved according to a network of chemical reaction rates and heating/cooling rates (\citealt{glover&abel2008}).  This network includes the relevant cooling and heating mechanisms for primordial gas in the density and temperature ranges covered in our simulation: H$_2$ collisional cooling, H and He collisional excitation and ionization, recombination, bremsstrahlung, inverse Compton scattering, etc.

It is worth further noting some of the processes that become relevant specifically at high densities ($n \ga 10^8$ cm$^{-3}$).
First, the chemothermal network includes three-body H$_2$ formation and H$_2$ formation heating.  
The gas evolution can be sensitive to the choice of various published three-body H$_2$ formation rates (\citealt{turketal2011}), so we use the intermediate rates presented in \cite{pallaetal1983}.  
Our network additionally includes the appropriate rates for H$_2$ photodissociation heating and UV pumping (e.g. \citealt{draine&bertoldi1996}).

At higher densities of $n \ga 10^9$ cm$^{-3}$, the gas becomes optically thick to H$_2$ ro-vibrational lines.  This leads to an opacity-dependent reduction in the rate at which the gas cools through these emission lines.  The chemistry and cooling routine estimates the optical depth using an escape probability formalism in combination with the Sobolev approximation.  We refer the reader to \cite{yoshidaetal2006} and \cite{greifetal2011} for further details, and to \cite{greif2014} and \nocite{hartwigetal2015b} Hartwig et al. (2015b) for more recent discussion of accurately estimating optically thick H$_2$ cooling rates.

We finally note that the chemothermal network includes H$_2$ collision-induced emission (CIE) cooling, which becomes significant when $n \ga 10^{14}$ cm$^{-3}$ (\citealt{frommhold1994}).  We handle the CIE cooling in the same way as described in  \cite{greifetal2011} and \cite{stacy&bromm2014}, but see also \cite{hirano&yoshida2013} for further discussion.

\subsection{Sink Particle Method}

Our sink particle method has been previously described in, e.g., \cite{stacy&bromm2014}. We briefly reiterate the main aspects of the sink algorithm in this section.

When an SPH particle reaches a density of  $n_{\rm max} = 10^{16}$\,cm$^{-3}$, its Jeans mass is approaching the resolution mass of the simulation, $M_{\rm J} \sim M_{\rm res} \simeq 3 \times 10^{-2}$ M$_{\odot}$.  At this point the simulation cannot be continued to higher densities since $M_{\rm J}$ would no longer be resolved.  Particles that condense to $n_{\rm max}$ are instead converted to sink particles.
Sink particles are described by an accretion radius  $r_{\rm acc}$, which we set equal to the resolution length such that
$r_{\rm acc} = L_{\rm res} \simeq 1.0$ AU, where

\begin{equation}
L_{\rm res}\simeq \left(\frac{M_{\rm res}}{\rho_{\rm max}}\right)^{1/3} \mbox{\ .}
\end{equation}

\noindent The sink accretes a nearby SPH particle if it is within distance $d < r_{\rm acc}$ and if it is not rotationally supported against infall onto the sink ($j_{\rm SPH} < j_{\rm cent}$, where $j_{\rm SPH}$ is the particle's specific angular momentum 
 with respect to the sink,
and $j_{\rm cent} = \sqrt{G M_*  d}$).  
 Effects of gaseous friction cannot be resolved on scales below $r_{\rm acc}$, but angular momentum may be removed on these unresolved scales.  We account for this effect by additionally accreting all particles with $d < 0.5 \, r_{\rm acc}$, regardless of the value of $j_{\rm SPH}$.

Accreted particles are removed from the simulation, and their mass is added to that of the sink.  When a sink first forms it immediately accretes most of the particles within $r_{\rm acc}$, giving the sink an initial mass of $M_{\rm res} \simeq  3 \times 10^{-2}$ M$_{\odot}$.  After each accretion event, the sink's position and velocity are set to the mass-weighted average of that of the sink and the accreted particles.  The sink is held at a constant density $10^{16}$ cm$^{-3}$ and temperature of 2000 K, the typical temperature for gas at this density.  The sink thus exerts a pressure on the surrounding particles, avoiding the formation of an artificial pressure vacuum around its accretion radius (see \citealt{bateetal1995,brommetal2002,marteletal2006}). 

Our algorithm also allows for two sinks to merge.  A larger sink and secondary sink are merged if their relative distance is 
$d < 0.5 \, r_{\rm acc}$, or if $d < r_{\rm acc}$ and $j_{\rm sec} < j_{\rm cent}$, where $j_{\rm sec}$ is the specific angular momentum of the secondary sink.  We note that many sink algorithms used in other work do not allow merging (e.g.  \citealt{smithetal2012, susaetal2014}).  Different merging algorithms, or not allowing sink mergers at all, may alter the sink accretion history  (e.g. \citealt{greifetal2011, shraderetal2014}).  
However, the most highly refined simulations of primordial discs find that on the small scales we simulate here, protostellar mergers will be quite common.  For instance,  \cite{greifetal2012} resolved sub-protostellar scales (0.05 R$_{\odot}$) and followed mergers without the use of sinks, finding that roughly half of the protostars formed will eventually be lost to mergers.
In this current work we find a similar rate of merger events (see Section 4 for further discussion). 
Thus, our merging algorithm may reflect what occurs on sub-sink scales. 
However, we stress that \cite{greifetal2012} followed their simulation for only the first 10 yr after initial protostar formation.  Determining the exact merger rate at later times still requires resolving sub-protostellar scales, which is not numerically feasible for our simulation timescales of several thousand years.

Our sink algorithm allows us to follow protostar formation and growth and protostellar merging events on a 1~AU scale over thousands of free-fall times ($\sim$\,10$^4$ yr).  Through sinks we avoid the computational cost of going to smaller resolution scales and even higher densities, all while gaining an unprecedented new look at the build-up of a Pop~III IMF.

\subsection{Ray-tracing Scheme}

We model the radiative feedback from the most massive protostar in our simulation, 
represented by the first sink that forms, which is also the most massive sink throughout the simulation.
We employ the same ray-tracing scheme described in detail in  \cite{greifetal2009} and \cite{stacyetal2012}. 
Here we provide a concise summary of the ray-tracer.

The largest sink particle (i.e., the most massive protostar) serves as a point source for photodissociating Lyman-Werner (LW) radiation as well as ionizing radiation.  
We note that radiative output is followed for only the most massive sink, while we do not account for the effect of additional feedback from secondary sinks.  This is distinct from previous work by, e.g., \cite{susaetal2014} who followed photodissociating feedback from all protostars but did not follow the propagation of the ionization front.

The rate of the ionizing emission $\dot{N}_{\rm ion}$ 
 from the largest sink
 is determined according to the protostellar evolution model described in Section 3 and assuming a blackbody spectrum.  The ray-tracer uses the sink position and $\dot{N}_{\rm ion}$ as its initial inputs and generates a spherical grid centred around the sink.  The grid consists of $\sim$\,10$^5$ rays and 200 logarithmically spaced radial bins.  The smallest radial bin is set to the distance between the sink and the nearest neighbouring SPH particle (typically $\sim$\,1 AU), while the maximum radius is set to the size of the entire cut-out region (3 pc).  The rays and radial bins total to $2\times10^{7}$ cells.  Each cell is characterized by a density and a set of chemical abundances, determined by the density-squared average of the SPH particles whose smoothing kernels overlap with the cell.  

The ray-tracer is updated every 5 hydrodynamic timesteps,
$t_{\rm hydro}$, 
which is sufficiently frequent since $t_{\rm hydro}$ is generally less than one-tenth of the sound-crossing time through the SPH kernel.  
Upon each update the grid is re-generated, and the most recent $\dot{N}_{\rm ion}$ from the protostellar evolution model is used to solve the I-front equation along each ray in the same manner as described in \cite{greifetal2009,stacyetal2012}.  Particles found to be within the  H~{\sc ii} region are given extra heating and ionization rates which are determined by the protostellar evolution model and distance from the sink, and these rates are passed into the chemothermal network.

Ionized particles around the sink range in hydrogen number density from $\sim n_{\rm H} \sim 10^{12}$ to $10^{9}$ cm$^{-3}$ (see Section 5.2), 
and their corresponding smoothing lengths can be described as follows:
\begin{equation}
h \simeq 0.5 {\rm AU} \left( \frac{  10^{16} \rm{cm^{-3}}  } {   n_{\rm H}  } \right)^{1/3}  \sim  10^{14} - 10^{16} \, {\rm cm} \mbox{,}  
\end{equation} 
or $ \sim 10 - 100 \, {\rm AU} $.  In comparison, for an ionizing emission rate of  $\dot{N}_{\rm ion} \sim 10^{48}$ s$^{-1}$, the Stromgren radius is
\begin{equation}
R_s = \left( \frac{3 \dot{N}_{\rm ion}} {4 \pi \alpha_{\rm B} n_{\rm HI}^2}  \right)^{1/3} \simeq 10^{12} -10^{14} \, {\rm cm}
\end{equation} 
As discussed in, e.g., \cite{susa2013}, at these high densities the SPH particles cannot resolve the initial Stromgren radius.  
When the protostar's effective temperature reaches $10^4$ K, we instead drive the initial expansion using a simple sub-grid model.  When the ray tracer solves the Stromgren equation within the central 200 AU, instead of using the true SPH-derived density $n_{\rm SPH}$, it uses a density $n_{\rm ch}$ based upon the analytic self-similar champagne flow solution (\citealt{shuetal2002}).  In this model, the H{\sc ii} region expands into an isothermal core as the central density $n_{\rm ch}$ declines.  The ray tracer uses the value $n_{\rm ch}$ until it declines to  $n_{\rm ch,min} = 10^7$ cm$^{-3}$, after which the ray tracer uses whichever is the smaller value between $n_{\rm SPH}$ and $n_{\rm ch,min}$.
While this method allows the central few hundred AU around the sink to become ionized at the expected time, self-consistently following the I-front breakout at these scales would require greater resolution than numerically feasible for our work.

In addition, previous work (e.g. \citealt{mckee&tan2008,hosokawaetal2011, stacyetal2012}) shows that I-fronts will first expand along the less dense polar regions while the dense disc remains shielded from radiation.  We assume that gas along 95\% of the rays is shielded from ionizing radiation, while the I-front expands only along the 5\% of radial directions which contain the lowest densities in the vicinity of the sink.

While the sink develops an H~{\sc ii} region, it never reaches sufficiently high effective temperature to develop an He~{\sc iii} region.  Though the ray-tracer is capable of following the growth of both H~{\sc ii} and He~{\sc iii}, in this case the `outer boundary' of the He~{\sc iii} region remains at zero (i.e., the location of the sink).  This in turn serves as the `inner boundary' of the H~{\sc ii} region.  We also note that for simplicity the ray-tracer does not distinguish between the H~{\sc ii} and He~{\sc ii} regions. 

The ray-tracer also is used to sum the H$_2$ column density $N_{\rm H_2}$ along each ray.   These $N_{\rm H_2}$ values are correspondingly assigned to each SPH particle, 
where each particle inherits the $N_{\rm H_2}$  value from the grid cell in which it sits.
The value is then
used to calculate the shielding factor $f_{\rm shield}$ (\citealt{draine&bertoldi1996}, but see also \citealt{wolcottetal2011,wolcott&haiman2011}).  Combined with the protostellar luminosity, $f_{\rm shield}$ is then used to calculate the H$_2$ dissociation rate (\citealt{abeletal1997}).  Unlike the ionization rates which are applied only to the  H~{\sc ii}-region particles, photodissociation rates are applied to every SPH particle.  However, most of the particles have sufficiently large shielding factors to make the photodissociation rates negligible.

We note that instead of applying the same value of $N_{\rm H_2}$ to each SPH particle of a grid cell, it will be more accurate in future work to apply an interpolation method to ensure LW photon conservation as the rays pass through each cell.  Our current method likely underestimates the strength of LW feedback in the disc, though we nevertheless see significant LW effects in the latter part of our simulation (see Section 5.1).  We finally note that we do not include effects of accretion luminosity heating of the gas, and this should also be included in future work.  This may reduce fragmentation at early times, but is unlikely to prevent the eventual development of vigorous fragmentation (e.g. \citealt{smithetal2011}).  Furthermore, in the latter part of the simulation UV luminosity effects dominate over accretion luminosity, so accretion luminosity becomes only a secondary effect.

\subsection{Radiation Pressure}

As the most massive sink, also referred to as the `main' sink,
grows in mass, 
the protostar it represents evolves in mass $M_*$, luminosity $L_*$, and rate of ionizing photon output  $\dot{N}_{\rm ion}$ .  We will discuss the protostellar evolution in more detail in Section 3.  
In this section we estimate the resulting effects of radiation pressure on the accreting gas for a given mass and luminosity and describe how we model these effects in the simulation.

In \cite{stacyetal2012} we presented estimates for effects of radiation pressure from Thomson and  Ly$\alpha$ scattering as well as ionizing radiation, and we briefly revisit those estimates here.
Thomson scattering radiation pressure is estimated to be dynamically important and able to drive an outflow when the source luminosity is greater than the Eddington luminosity: 
\begin{equation}
L_* \ga L_{\rm Edd} = 4\pi G M_* m_{\rm H} c/\sigma_{\rm T} \mbox{,}
\end{equation} 
where $M_*$ is the stellar mass, $m_{\rm H}$ the mass of a hydrogen atom, 
and $\sigma_{\rm T}$ the Thomson scattering cross section.  
For $M_* = 20$ M$_{\odot}$, the expected stellar luminosity is $L_*\sim 10^5$ L$_{\odot}$ while 
$L_{\rm Edd} \simeq 6 \times 10^5$ L$_{\odot}$.  
For  $M_* \sim 100$ M$_{\rm \odot}$, we instead have $L_* \sim L_{\rm Edd} \sim 3\times10^6$ L$_{\odot}$, 
depending on the  
protostellar radius
and accretion rate onto the protostar (e.g. \citealt{omukai&palla2003}).
Thus, Thomson scattering becomes effective only for more extreme masses.
However, in our simulation the main protostar does not grow above 20 M$_{\odot}$.
We therefore do not include the effects of Thomson scattering in our simulation.

An approximation for pressure from Ly$\alpha$ can be found in, e.g., \cite{bithell1990, oh&haiman2002}, and was also presented in \cite{stacyetal2012}:

\begin{equation}
P_\alpha = \frac{\dot{N}_{\rm Ly\alpha}}{4\pi r^2} \left(\frac{h \nu_{\alpha}}{c} \right) N_{\rm scat} \mbox{\ ,}
\end{equation}

\noindent where $\dot{N}_{\rm Ly\alpha}$ is the emission rate of Ly$\alpha$ photons,  $h\nu_{\alpha} \sim 10.2$~eV is the energy of a Ly$\alpha$ photon,  and 
$N_{\rm scat}$ is the number of scatterings per photon before it exits the region in question
(see also the discussion and more detailed formulation in \citealt{mckee&tan2008} and \citealt{mbco2009}).  
Consider a newly formed H~{\sc ii} region that has reached a size of 10 AU.  
 From the expressions in the above-mentioned studies,
 we estimate $N_{\rm scat}\simeq 100$
 for the ionized region and one to two orders of magnitude higher in a similarly sized neutral region
 (e.g. \citealt{bithell1990,haehnelt1995,mckee&tan2008}).  
 Given $\dot{N}_{\rm Ly\alpha} \sim \dot{N}_{\rm ion} \sim 5 \times 10^{48}$ s$^{-1}$ for a $\sim$ 20  M$_{\odot}$ protostar, we find upper limits to $P_\alpha$ of $\sim 0.1$ dyn cm$^{-2}$ within the H~{\sc ii}~region.  
 In comparison, the gas has density and temperature of approximately 10$^{13}$ cm$^{-3}$ and 10$^4$ K, yielding a gas pressure of
 $P_{\rm therm} \sim nk_{\rm B}T \sim 10$ dyn cm$^{-2}$.  
 Within the  H~{\sc ii}~region, thermal pressure therefore dominates over Ly$\alpha$  pressure.  
 
The effect of $P_\alpha$ is more important for the shell of neutral gas surrounding the  H~{\sc ii} region, which is much more optically thick to Ly$\alpha$ photons (e.g. \citealt{mbco2009}). 
 Indeed, as estimated in \cite{mckee&tan2008} and \cite{stacyetal2012}, Ly$\alpha$  pressure can be large enough to overcome the ram pressure of infalling gas at the edge of a 10,000~AU H~{\sc ii}~region when the protostar has reached $\sim$\,20 M$_{\odot}$.   
 However, because Ly$\alpha$ 
 photons
 will travel much more freely through the diffuse and ionized polar regions than through the disc, 
 break-out should occur at the poles first.
This alleviates 
the radiation
pressure closer to the disc plane, so the protostellar accretion disc should remain unaffected.  
We thus do not expect that inclusion of Ly$\alpha$ pressure would significantly change our results, and we do not include it in our model.

Pressure from ionizing radiation can become similar in importance to gravity and gas pressure on certain scales (see also \citealt{haehnelt1995,mckee&tan2008,stacyetal2012}). 
Again estimating $\dot{N_{\rm ion}} = 5 \times 10^{48}$ s$^{-1}$, 
radiation pressure within an H~{\sc ii} region dominates gravity on scales of $\sim$\,100 AU and beyond (see equation 48 of \citealt{mckee&tan2008}).  
However, this scale is similar to that of the gravitational radius,
the radius at which $P_{\rm therm}$ in turn dominates over gravity, and the sound speed of the ionized gas exceeds the free-fall velocity.  
The gravitational radius is given by

\begin{equation}
r_{\rm g} \sim \frac{G M_*} {c_s^2} \sim 200 {\rm AU} \mbox{,}
\end{equation}

\noindent where $M_* \sim 20$ M$_{\odot}$ and $c_{\rm s} \sim 10$ km s$^{-1}$.

The force of diffuse ionizing radiation can be approximated by 

\begin{equation}
f_{\rm ion,diff} = \alpha_{\rm 1} x_{\rm ion}^2 n_{\rm H} \left(\frac{h\nu_i}{c}\right)
\end{equation}

\noindent  (c.f. \citealt{haehnelt1995}, equation 3).  
We set $h\nu_i = 13.6$ eV,  while 
$x_{\rm ion}$  is the ionization fraction, and
 $\alpha_{\rm 1}$ is the recombination coefficient to the ground state ($\simeq 1.6 \times 10^{-13}$ cm$^{3}$s$^{-1}$; \citealt{osterbrock&ferland2006}).
We approximate the corresponding diffuse ionizing radiation pressure on the gas represented by an SPH particle as

 \begin{equation}
 P_{\rm ion,diff} \sim \alpha_{\rm 1} x_{\rm ion}^2 n_{\rm H}  \left(\frac{h\nu_i}{c}\right) \left( \frac{1}{m_{\rm H}} \right) \left( \frac{m_{\rm SPH}}{h_i^2} \right)
 \mbox{\ ,}
 \end{equation}

\noindent where  $P_{\rm ion,diff}$  represents the diffuse 
pressure from re-emitted ionizing photons generated by  recombinations to the ground state 
(e.g. \citealt{omukai&inutsuka2002,hosokawaetal2011}), 
and  $h_i$ the smoothing length of the SPH particle. 
This formulation applies the `on-the-spot' approximation, which assumes that each photon is reabsorbed very close to the point at which it was generated (\citealt{osterbrock&ferland2006}). 
This is an appropriate assumption, particularly when comparing the photon scattering length to the size of the H~{\sc ii} region as well as the scale of the particle smoothing lengths and I-front radial bins.

For gas within the   H~{\sc ii} region we account for diffuse ionization pressure by adding $P_{\rm ion,diff}$ to the thermal pressure term in the calculation of hydrodynamic acceleration (see \citealt{springel2005}).
For each particle we then set its total pressure to $P = P_{\rm therm} + P_{\rm ion,diff}$.

In addition, ionized particles receive an additional acceleration term from direct ionization using the following:

 \begin{equation}
 a_{\rm ion,dir}  \sim   \frac{k_{{\rm HI}}n_{{\rm HI}}}{\rho_{\rm H}}  \left(\frac{h\nu_i}{c}\right) 
 =  \frac{k_{{\rm HI}}\left(1- x_{\rm ion}\right)}{m_{\rm H}}  \left(\frac{h\nu_i}{c}\right)
 \mbox{\ ,}
 \end{equation}

\noindent where $k_{{\rm HI}}$ is the photoionization rate of  H{\sc i},    
$n_{{\rm HI}}$ is the H{\sc i} number density, 
and $\rho_{\rm H}$ is the total hydrogen mass density.
(e.g. \citealt{johnsonetal2011, park&ricotti2012}; see also \citealt{hosokawaetal2011}).
In contrast to the diffuse radiation, the force of direct ionization acts preferentially in the direction radially away from the emitting sink. We thus add $a_{\rm ion,dir}$ to the sum of the gravitational and hydrodynamical accelerations accordingly, resulting in a velocity kick in the direction opposite to the sink (e.g. \citealt{johnsonetal2011}).

\begin{figure*}
\includegraphics[width=.46\textwidth]{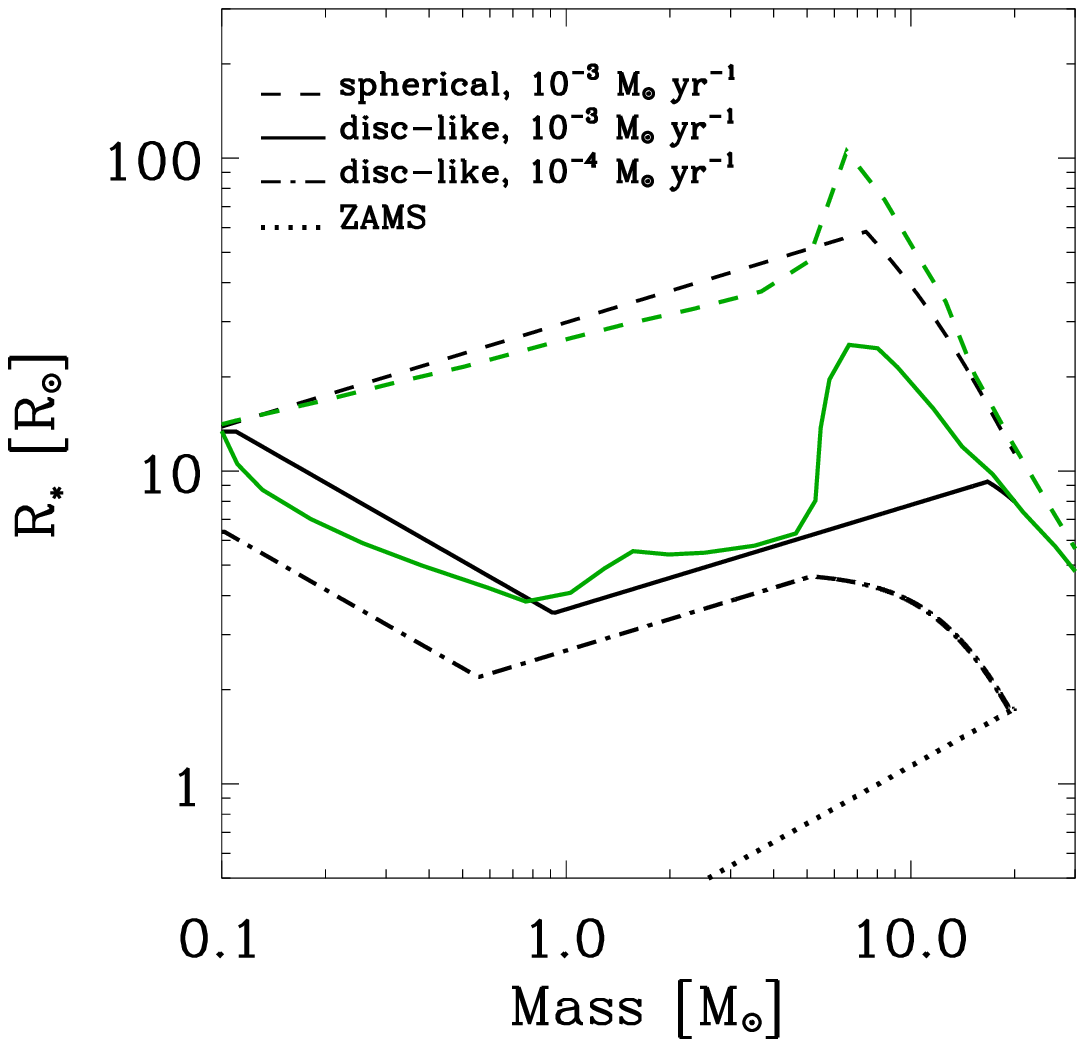}
\includegraphics[width=.46\textwidth]{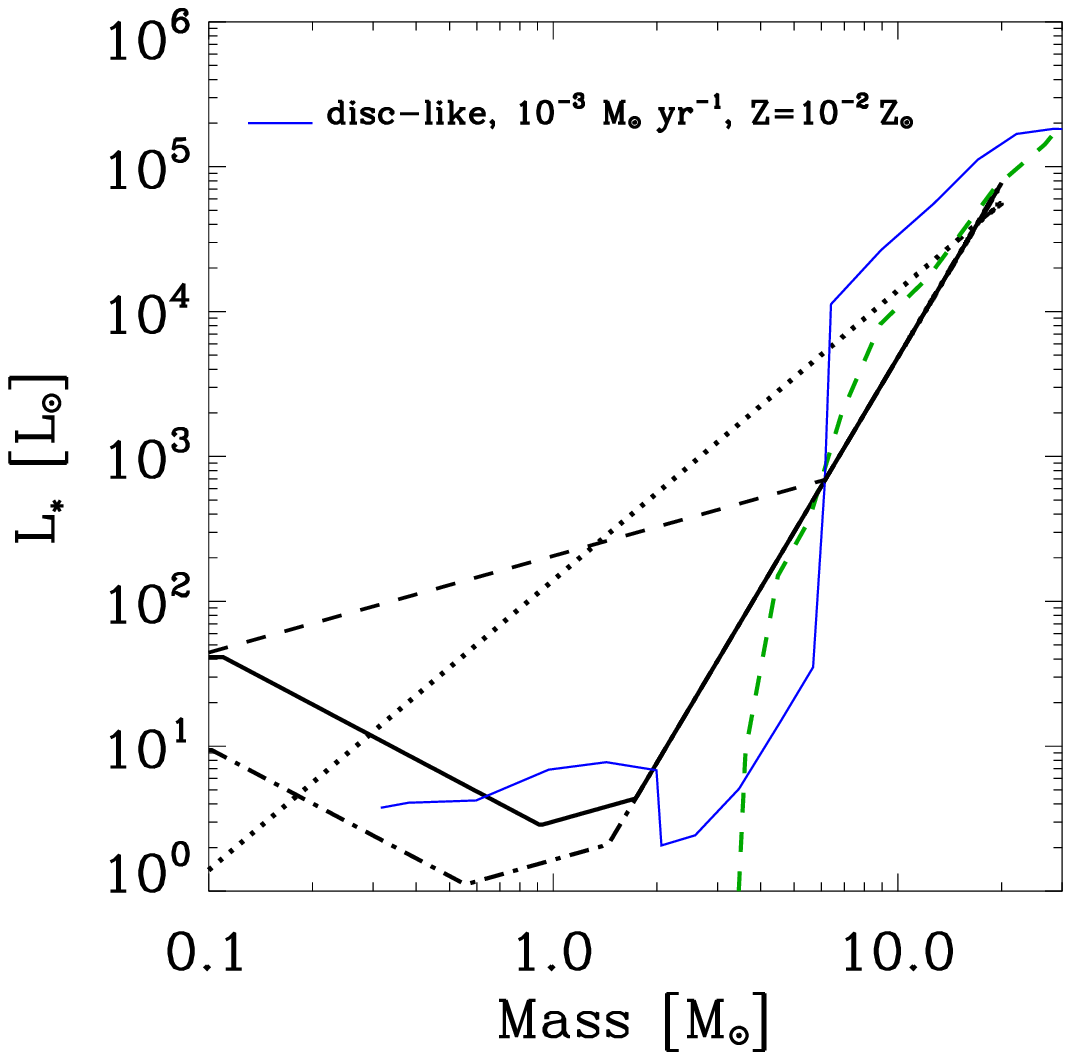}
 \caption{
{\it Left panel:}  Protostellar radius versus mass given a constant accretion rate of $\dot{M} = 10^{-3}$ M$_{\odot}$ yr$^{-1}$ (compare with fig. 11 of Hosokawa et al. 2010).  
Green lines are taken directly from Hosokawa et al. (2010).  Black lines represent the corresponding radial evolution as taken from the models employed in our numerical simulation.  Dashed lines represent the radial evolution for the `spherical accretion' mode, and solid lines show the evolution for `spherical-to-disc' accretion, where the accretion geometry is switched at $M_* = 0.1$ M$_{\odot}$.  The dash-dotted line shows evolution within our model for  $\dot{M} = 10^{-4}$ M$_{\odot}$ yr$^{-1}$ in the same `spherical-to-disc' accretion scenario.  The dotted line shows $R_{\rm ZAMS}$ for a primordial protostar.  Our model well-matches the more detailed calculations of Hosokawa et al. (2010) for a range of geometries and accretion rates.  The `spherical' accretion models go through an initial adiabatic expansion phase, while the `spherical-to-disc' models first undergo a radial decline until deuterium burning increases the average entropy in the protostar and allows it to expand.  The fits to the Hosokawa et al. (2010) results furthermore show more rapid expansion during the `swelling' phase at $\la$  10 $M_{\odot}$, not included in our model.  The protostars subsequently undergo KH contraction to the MS.  
{\it Right panel:}  Protostellar luminosities for the same set of models.  For comparison to disc evolution at a larger metallicity, solid blue line is from fig. 5 of Hosokawa et al. (2010), which shows luminosity evolution for metallicity of 0.02 Z$_{\odot}$.  Dashed green line is from figure 4 of Omukai and Palla (2003).  Note the general convergence of the luminosities at $M_* > 5$ M$_{\odot}$. 
}
\label{star-rad}
\end{figure*}

\section{Protostellar Evolution Model}

As we model the protostellar feedback
from the first-forming and most massive sink (the `main' sink), 
each update of the ray-tracer requires an input of the current ionizing photon rate,
 $\dot{N}_{\rm ion}$, which is determined by the protostellar effective temperature $T_{\rm eff}$ and luminosity $L_*$.   These are in turn determined by the protostellar mass  $M_*$, the history of the accretion rate $\dot{M}$, and the protostellar  radius $R_*$.  Instead of directly solving equations of stellar structure, we approximate the protostellar evolution using analytic fits to the results of previous studies (e.g. \citealt{stahler&palla1986,omukai&palla2003, hosokawaetal2010}).  This was briefly described in \cite{stacy&bromm2014}, but we provide more detail below. 

\subsection{Luminosity and Temperature Evolution}

The total luminosity $L_*$ is determined by two components, the accretion luminosity $L_{\rm acc}$ and the internal luminosity $L_{\rm int}$.  $L_{\rm acc}$ originates from energy released at the accretion shock on the surface of the star, while $L_{\rm int}$ originates from energy generated in the stellar interior.  We thus have

\begin{eqnarray}
L_* &=& L_{\rm acc} + L_{\rm int} 
= \alpha  \frac{G M_* \dot{ M}}{R_*} 
+  L_{\rm int} \mbox{,}
\end{eqnarray}

\noindent (e.g. \citealt{prialnik&livio1985,hartmannetal1997}).  
We set $M_*$ to be the mass of the main sink.  
Sink accretion of SPH particles occurs in discrete events, so we smooth the accretion rate $\dot{M}$ over time intervals of 10 yr to estimate a true physical accretion rate (i.e. $\dot{M} = [M_*(t) - M_*(t-10 {\rm yr})] / 10 {\rm yr}$).  
10 yr is over two orders of magnitude longer than the shortest $t_{\rm hydro}$, but short enough to resolve minimal accretion rates of 
$< 10^{-5}$ M$_{\odot}$ yr$^{-1}$.
If the sink has accreted no particles within the last 10 yr, then $\dot{M} = 0$ and $L_* = L_{\rm int}$.

The amount of energy released at the accretion shock depends upon
the details of the mass flow as it approaches the stellar surface,
and this is parameterized by $\alpha$.  For 
the extreme limiting case of pure
cold disc accretion in which gas smoothly settles onto the star, $\alpha=0$.  For 
the opposite limit of
hot spherical accretion in which the entire star is surrounded by an accretion shock, $\alpha=1$.  
As in \cite{stacy&bromm2014}, we estimate $\alpha$ by first measuring the specific angular momentum $j_{\rm SPH}$ of each particle accreted by the main sink within the last 10 yr, as well as each particle currently within 10 AU from the sink.  
$N$ is the total number of such gas particles.  Slightly different from \cite{stacy&bromm2014}, however, we define $\alpha$ as

\begin{equation}
\alpha = 1 - \frac{1}{N}  \sum\limits_{i=0}^N { \frac{j_{\rm SPH,i}}{j_{\rm cent ,i}} } \mbox{.}
\end{equation}

\noindent Thus, 
we assume
$\alpha \sim 1$ if the near-sink gas is is dominated by radial motion, while $\alpha \sim 0$ if it is dominated by rotational motion.
While this allows us to assign a value to the amount of entropy the accreting mass carries into the stellar interior, we cannot determine the true value at our given resolution.  Furthermore, even with disc accretion some amount of entropy may still be advected into the star, particularly for rapid accretion rates (e.g. \citealt{hosokawaetal2013}).

When the main sink first forms, we determine $L_{\rm int}$ by assuming the protostar is born on the Hayashi track of the Hertzsprung-Russell diagram.  As discussed in, e.g., \cite{stahleretal1986} and mentioned in \cite{stacy&bromm2014}, Hayashi tracks are described by a constant effective temperature which can range from $\sim$\,3000 to 5000 K depending upon the stellar mass and opacity.   $T_{\rm Hay}$  tends to be higher for the metal-free atmospheres of Pop III stars, so we choose a value in the upper end of this range, $T_{\rm Hay} = 4500$ K.  The `Hayashi track' luminosity is then

\begin{eqnarray}
L_{\rm Hay} = 4 \pi R_*^2 \sigma_{\rm SB} T_{\rm Hay}^4 \mbox{.}
\end{eqnarray}

\noindent As the protostar continues to grow in mass it eventually transitions to the Henyey track.  A protostar of a given mass on the Henyey track has a relatively constant $L_{\rm int}$ while $T_{\rm eff}$ and $R_*$ will vary.  The `Henyey track' luminosity can be fitted to the mass and effective temperature using

\begin{eqnarray}
L_{\rm Hen} =   10^4 {\rm L_{\odot}} \left(\frac{M_*}{9{\rm M_{\odot}}}\right)^{22/5} \left(\frac{T_{\rm eff}}{10^4 \rm K}\right)^{4/5}  \mbox{.}
\end{eqnarray}

\noindent (\citealt{henyeyetal1955}, see also \citealt{hansenetal2004}).  We note that, particularly at low masses, $L_{\rm Hen}$ will tend to be larger for zero-metallicity stars than for corresponding stars of higher metallicity (e.g. \citealt{stahleretal1986,cassisi&castellani1993}), again due to differences in opacity.  We have thus chosen a fit for $L_{\rm Hen}$ that is most appropriate for a typical Pop III protostellar mass of $\sim$\,10 M$_{\odot}$.
 
For the initial protostellar phases, we first set $L_{\rm int} = L_{\rm Hay}$, similar to the prescription described in \cite{offneretal2009}.  Once $L_{\rm Hen}$ grows larger than $L_{\rm Hay}$, we use $L_{\rm int} = L_{\rm Hen}$ until the protostar has contracted down to the main-sequence radius.  After that point we set $L_{\rm int} = L_{\rm ZAMS}$, where  $L_{\rm ZAMS}$ is the  zero-age main sequence (ZAMS) luminosity of the protostar once it has reached the onset of hydrogen burning.  We can describe $L_{\rm ZAMS}$ using a simple fit to stellar mass:

\begin{equation}
L_{\rm ZAMS} = 1.4 \times 10^4 {\rm L_{\odot}} \left(\frac{M_*}{10 \rm M_{\odot}}\right)^2 \mbox{ }
\end{equation}

\noindent (e.g. \citealt{schalleretal1992,hosokawaetal2010}).  

\subsection{Radial Evolution}

The radial evolution of a protostar growing with a constant accretion rate can be described by two main phases - the `adiabatic accretion' phase followed by the Kelvin-Helmholtz (KH) contraction phase (\citealt{stahler&palla1986,omukai&palla2003,hosokawaetal2010}).  When the protostar first forms, the accretion timescale $t_{\rm acc}$ is significantly shorter than the KH timescale  $t_{\rm KH}$ on which the star can radiate energy and contract.  These timescales can be written as
 
\begin{equation}
t_{\rm KH} = \frac{G M_*^2}{R_*L_*}  
\end{equation}

\noindent and

\begin{equation}
t_{\rm acc} = \frac{M_*}{\dot{M}} \mbox{.}
\end{equation}

\noindent   During the initial adiabatic accretion phase, $R_*$ gradually increases with mass.    
As the star grows, $t_{\rm KH}$ declines while  $t_{\rm acc}$ steadily increases until $t_{\rm acc} \ga  t_{\rm KH}$, at which point KH contraction commences.
 
As described in, e.g., \cite{hosokawaetal2010}, the radial evolution has important dependencies upon the accretion geometry.  
Protostars undergoing geometrically thin disc accretion have smaller radii than spherically symmetric accretion.
As mentioned above, spherical accretion leads to an accretion shock around the entire stellar surface.
In contrast, perfectly thin disc accretion adds minimal entropy to the protostar since most of the inflow is unlinked to the surface of the star.
The physical reality, however, is that the accretion geometry fluctuates between these two idealized cases.

We note that the semi-analytic model of \cite{tan&mckee2004} found radial evolution in the spherical case very similar to that of \cite{hosokawaetal2010}.  However, their disc-like accretion model also yielded values of $R_*$ similar to those in the spherical case, since they still assume some thickness to the disc which allows for the action of an accretion shock upon the stellar surface.    
In our simulation the radial evolution is ultimately most similar to the spherical case, and thus closely follows the results of both \cite{hosokawaetal2010} and \cite{tan&mckee2004}.

Below we describe the radial evolution of the protostar in the `spherical' case, in which the protostar gains the entirety of its mass through spherical accretion (Section 3.2.1).  We also describe the `spherical-to-disc' case, in which the protostar initially grows through spherical accretion but later transitions to accretion through a disc (section 3.2.2).  Though the models we present describe the entire evolution of the protostar, we note that not all of these evolutionary phases are actually utilized in our simulation.  In particular, our simulation does not follow the protostellar evolution to the point of KH contraction to the MS.  However, we include descriptions of these phases for completeness.

\subsubsection{Radial Evolution under Spherical Accretion}

The radial evolution during the adiabatic accretion phase for spherical accretion can be described as

\begin{equation}
R_{I,\rm sphere} \simeq 49 {\rm R_{\odot}} \left(\frac{M_*}{\rm M_{\odot}}\right)^{1/3} \left(\frac{\dot{M}}{\dot{M}_{\rm fid}}\right)^{1/3}   \mbox{\ .}
\end{equation}
      
\noindent (\citealt{stahler&palla1986, omukai&palla2003}).
To model the subsequent KH contraction we use the same fit as in 
\cite{stacyetal2010} and \cite{stacyetal2012}, which was based on the above-cited studies of spherical accretion:

\begin{equation}
R_{II,\rm sphere} \simeq 140 {\rm R_{\odot}} \left(\frac{\dot{M}}{\dot{M}_{\rm fid}}\right) \left(\frac{M_*}{10 \rm M_{\odot}}\right)^{-2} \mbox{\ ,}
\end{equation}

\noindent where  $\dot{M}_{\rm fid}\simeq 4.4\times 10^{-3}$~M$_{\odot}$\,yr$^{-1}$, the fiducial accretion rate.  We assume KH contraction begins when $R_{*II,\rm sphere} <  R_{*I,\rm sphere}$. 
This occurs when $M_* \sim 10$ M$_{\odot}$, in agreement with previous work (\citealt{omukai&palla2003,hosokawaetal2010}).

\subsubsection{Radial Evolution in Case of Transition to Disc Accretion}

If the accretion geometry transitions from spherical to disc-like, the radial evolution during the `adiabatic accretion' phase will diverge from the purely spherical case.  Power-law modeling of `spherical' and `spherical-to-disc' accretion, taken directly from the study of primordial stars growing at $\dot{M} = 10^{-3} \rm M_{\odot} yr^{-1}$ in Hosokawa et al. (2010; their fig. 11), is reproduced in Fig. \ref{star-rad}.  In their study, the transition in accretion geometry occurs for $M_* = 0.1$ M$_{\odot}$.  

For the range of accretion rates studied in \cite{hosokawaetal2010}, the radius rapidly declines due to the decrease in entropy imparted to the protostar, once the accretion geometry becomes disc-like.  We model this decline as 

\begin{equation}
R_{I,\rm disc} = R_{Ia,\rm disc} \simeq R_0 \left(\frac{M_*}{\rm M_0}\right)^{-0.63 }  \mbox{\ ,}
\end{equation}

\noindent where $M_0$ and $R_0$ are the protostellar mass and radius when the transition from spherical to disc-like accretion occurs.  

As the mass grows and radius declines, the interior temperature $T_{\rm int}$ evolves according to $T_{\rm int} \propto M_*/R_*$. More precisely, the models of \cite{hosokawaetal2010} find that $T_{\rm int}$ will increase as the protostar contracts according to

\begin{equation}
T_{\rm int} =  10^6 {\rm K} \left(\frac{M_*}{0.3 \rm M_{\odot}}\right) \left(\frac{R_*}{2.4 \rm R_{\odot}}\right)^{-1}  \mbox{\ .}
\end{equation}

\noindent The radial decline will continue until deuterium burning begins in the stellar interior and increases the average entropy within the star, which occurs when $T_{\rm int} > 2 \times 10^6$ K.  At this point the protostar again undergoes a roughly adiabatic expansion, with approximately the same powerlaw dependence on $M_*$ and $\dot{M}$ as found for spherical accretion:

\begin{equation}
R_{I,\rm disc} = R_{Ib,\rm disc} \simeq R_1 \left(\frac{M_*}{M_1}\right)^{1/3} \left(\frac{\dot{M}}{\dot{M_1}}\right)^{1/3}   \mbox{\ ,}
\end{equation}

\noindent where $R_1$, $M_1$, and $\dot{M_1}$ are the radius, mass, and accretion rate when deuterium burning commences.  

As found in \cite{omukai&palla2003} and \cite{hosokawaetal2010}, and explicitly quantified by \cite{smithetal2011}, for both disc and spherical accretion the protostar will begin its KH contraction phase at an approximate mass of 

\begin{equation}
M_2 \sim 7 {\rm M_{\odot}} \left(\frac{\dot{M}} {10^{-3} \rm M_{\odot} yr^{-1}}\right)^{0.27} \mbox{\ ,}
\end{equation}

\noindent roughly corresponding to the condition $t_{\rm acc} >  t_{\rm KH}$.  For a range of accretion rates, \cite{hosokawaetal2010} furthermore find that the maximum radius reached by the protostar undergoing purely spherical accretion is roughly three times larger than a protostar which transitions to disc accretion.  For disc accretion this maximum radius can be approximated by

\begin{equation}
R_2 \simeq 47 {\rm R_{\odot}} \left(\frac{\dot{M}}{\dot{M}_{\rm fid}}\right) \left(\frac{M_2}{10 \rm M_{\odot}}\right)^{-2}  \mbox{\ .}
\end{equation}

\noindent When this maximum radius is reached, KH contraction will begin.  
Once $M_2$ and $R_2$ are known, along with the mass and radius at which the protostar reaches the ZAMS  (see below), this can be used to derive the rate at which the protostar will KH contract onto the MS:

\begin{equation}
R_{II,\rm disc} \sim R_2 \left( \frac{M_*}{M_2} \right)^a  \mbox{\ ,}
\end{equation}
 
 \noindent where
 
\begin{equation}
a =  \frac{ {\rm log}\left(R_{\rm ZAMS} / R_2 \right)} { {\rm log}\left(M_{\rm ZAMS} / M_2\right)} \mbox{\ .}
\end{equation} 
 
 \noindent  We subsequently set our radius in between the spherical and disc cases such that prior to KH contraction

\begin{equation}
R_* = R_{I} = \alpha R_{I,\rm sphere} + \left( 1 - \alpha \right)R_{I,\rm disc} \mbox{\ ,}
\end{equation}

\noindent while after KH contraction

\begin{equation}
R_* = R_{II} = \alpha R_{II,\rm sphere} + \left( 1 - \alpha \right)R_{II,\rm disc} \mbox{\ .}
\end{equation}
 
 \begin{figure}
\includegraphics[width=.45\textwidth]{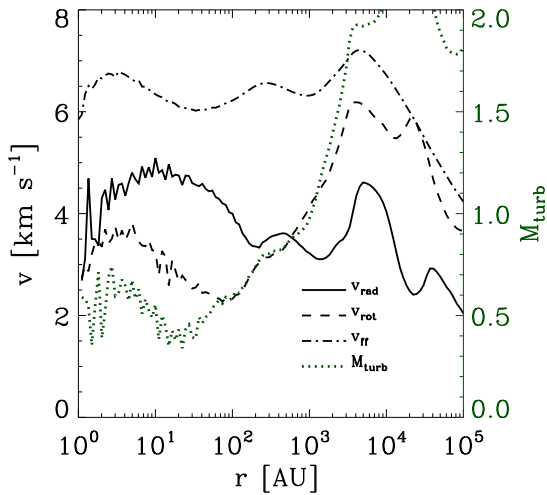}
\caption{
Velocity structure of gas just before initial sink formation.
Solid line is the magnitude of the radial infall velocity $v_{\rm rad}$, averaged within a range of logarithmic radial bins.  Dashed line is the average rotational velocity v$_{\rm rot}$ within these same bins.  
Dash-dotted line is the free-fall velocity v$_{\rm ff}$ based upon the enclosed mass at each radius. 
Dotted green line is the turbulent Mach number $M_{\rm turb}$.  
}
\label{velprof}
\end{figure}

\subsubsection{Radial Evolution Under Low and High Accretion Rates}

For very low accretion rates, the models described above can lead to unphysically small $R_*$, falling below the MS radius $R_{\rm ZAMS}$ or even becoming zero during periods of no accretion ($\dot{M}=0$).  In this case, the above equations cannot be used to obtain values for the protostellar radius.  This is equivalent to the protostellar radius unphysically declining more rapidly than the KH contraction rate.  When the radius declines rapidly enough that $\dot{R} < -6\,R_*/t_{\rm KH}$, we instead take  the last determined value for $R_*$ and assume the protostar declines in radius according to 

\begin{equation}
\dot{R} = \dot{R}_{\rm min} = -6 \, R_*/t_{\rm KH} \mbox{.} 
\end{equation}

This constraint is applied until either the MS radius is reached or $\dot{M}$ again becomes sufficiently large that 
 $\dot{R} \ge -6\,R_*/t_{\rm KH}$.
We choose our value of $\dot{R}_{\rm min}$ such that the protostar will smoothly decline in radius, but still contract rapidly enough to reach ionizing effective temperatures once it has grown to $\ga$ 10 M$_{\odot}$. 
The particular factor of 6 was both physically and numerically motivated.  We found this factor allowed the protostar to reach ionizing luminosities within a computationally feasible amount of time while still preventing the protostar from becoming ionizing before reaching a substantial mass of $\sim$ 20 M$_{\odot}$.
Previous models of e.g., \cite{omukai&palla2003} and \cite{smithetal2012}, show that under rapid spherical accretion a protostar may not become ionizing until reaching somewhat larger masses.  However, given the range of uncertainty in the models, our estimate falls in between the range of disc-like and spherical accretion cases while allowing the simulation to follow the growth of the I-front within a reasonable computation time.

Finally, we note that short bursts of rapid accretion may lead to very rapid protostellar expansion (e.g. \citealt{hosokawaetal2016}). 
At later times, when $M_* > 10$ M$_{\odot}$, 
we similarly impose a limit to the expansion rate, such that

\begin{equation}
\dot{R} \le \dot{R}_{\rm max} = R_*/t_{\rm KH} \mbox{.}
\end{equation}

\noindent Though this limit is somewhat artificial, the protostellar evolution is still within the range of uncertainty of the instantaneous accretion rate and the precise value for $\alpha$.  This also helped to ensure that once the protostar was well over 10 M$_{\odot}$ it would reach ionizing effective temperatures.  As we show in Section 4, it happens that the limit did not significantly alter the protostellar evolution in the latter part of the simulation.

\subsubsection{Approach to the Zero-Age Main Sequence}

Radial contraction will continue until the protostar reaches the ZAMS, which occurs at a radius of approximately 

\begin{equation}
R_{\rm ZAMS} =   0.28 {\rm R_{\odot}}  \left(\frac{M_*}{1 \rm M_{\odot}}\right)^{0.61}   \mbox{\ .}
\end{equation}

\noindent The corresponding luminosity at this point is

\begin{equation}
L_{\rm ZAMS} =   140 {\rm L_{\odot}}  \left(\frac{M_*}{1 \rm M_{\odot}}\right)^{2}   \mbox{\ .}
\end{equation}

\noindent \cite{hosokawaetal2010} have noted that $R_{\rm ZAMS}$ is significantly smaller for Pop III stars as compared with stars containing some metallicity.  This is due to the lack of C, N, and O nuclei in Pop III stars,  which delays the onset of CNO-cycle hydrogen burning until these elements are generated through helium triple-$\alpha$ burning.  Though the pp-chain is already active, this is not energetic enough to halt contraction, so the protostar shrinks to lower radii before reaching its ZAMS radius.  

Assuming a constant accretion rate, following the discussion in \cite{hosokawaetal2010}, we can estimate:
  
\begin{equation}  
M_{\rm ZAMS} \sim 50 {\rm M_{\odot}} \left(\frac{\dot{M}} {10^{-3} \rm M_{\odot} yr^{-1}}\right)^{0.62}  
\end{equation}
   
\noindent (\citealt{omukai&palla2003}).  Upon reaching the ZAMS, contraction halts, and we set  $L_{\rm int} = L_{\rm ZAMS}.$  

The above described models for `spherical' and `spherical-to-disc' accretion are shown in Fig. \ref{star-rad} given a constant $\dot{M} = 10^{-3} \rm M_{\odot} yr^{-1}$ and $10^{-4} \rm M_{\odot} yr^{-1}$, allowing for direct comparison to the full calculation done by \cite{hosokawaetal2010}.  Though for simplicity we do not include the `swelling' phase that occurs between adiabatic accretion and KH contraction, our model does well in reproducing the main phases of radial evolution found in their work.   

Note that the $\dot{R} \ge \dot{R}_{\rm min}$ constraint was applied for the lower accretion rate of  $\dot{M} = 10^{-4}$ M$_{\odot}$ yr$^{-1}$.  This can be seen in Fig. \ref{star-rad} where the $\dot{M} = 10^{-4}$ M$_{\odot}$ yr$^{-1}$ `spherical-to-disc' case declines to the ZAMS at the KH rate instead of the rate given by the function for $R_{II,\rm disc}$.

\begin{figure*}
\includegraphics[width=.85\textwidth]{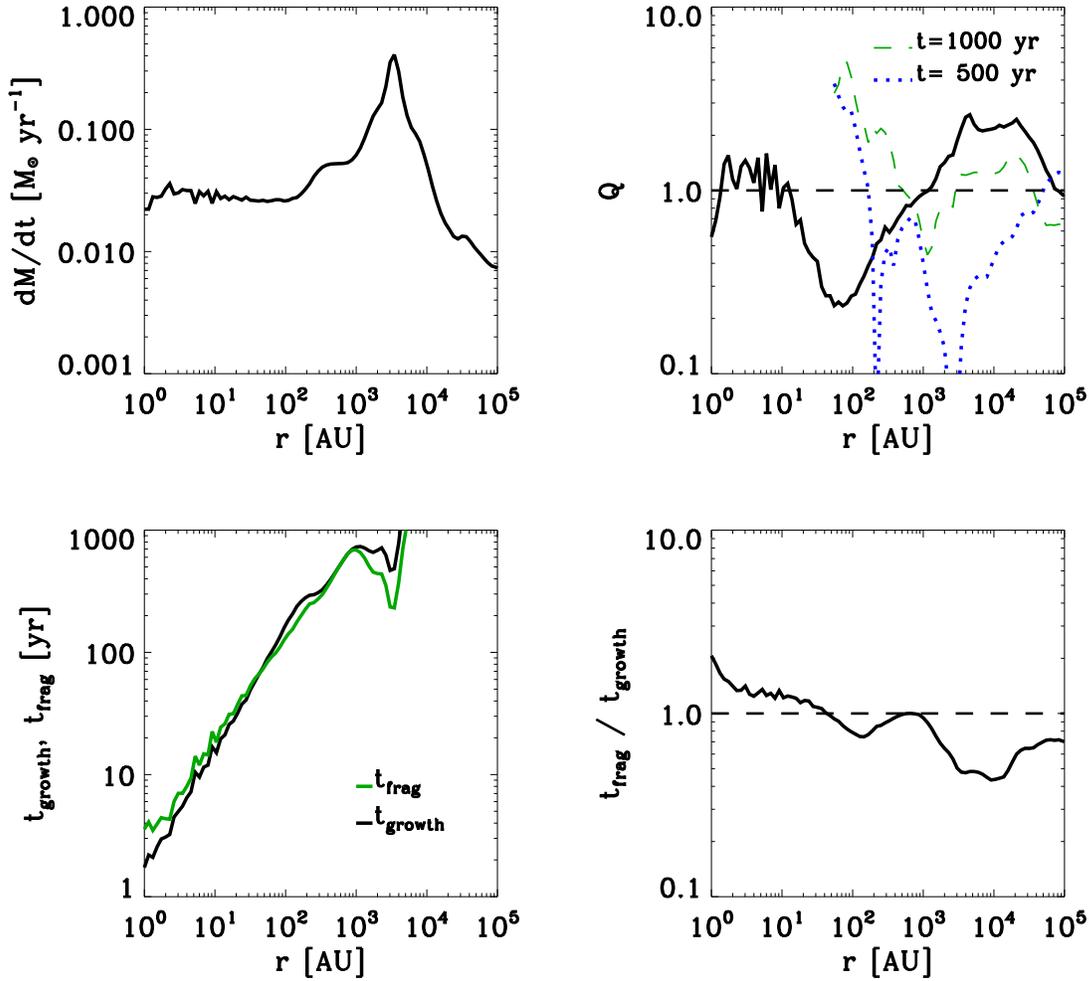}
 \caption{
 Radial profile of disc properties just prior to initial sink formation.
 Profiles are measured with respect to distance from the main sink.
 {\it Upper Left:} Spherical accretion rate estimate $\dot{M}_{\rm sphere}$.  
 In addition, dashed line shows ratio of rotational to gravitational energy.
 {\it Upper Right:}  Toomre parameter.  
 Dotted blue and dashed green lines additionally show $Q$ at 500 and 1000 yr.
 The horizontal dashed line represents $Q=1$, the value below which the Toomre criterion for fragmentation is satisfied.   
 {\it Lower Left:}   Growth timescale $t_{\rm growth}$ (black line) and fragmentation timescale $t_{\rm frag}$ (green line).
 {\it Lower Right:}  Ratio of $t_{\rm growth}$ to $t_{\rm frag}$, where values less than one indicate regions likely to undergo fragmentation.
 Note that as more mass is removed from the disc and onto the sinks, the surface density declines and the $Q$ parameter diverges at progressively larger radii.  Thus, the value of $Q$ at small radii has physical meaning only at early times.
 }
\label{toomre}
\end{figure*}


\section{Early Gas Evolution}

\subsection{Initial Minihalo Collapse}

The chemical and thermal evolution of the central minihalo gas up to the formation of the first sink particle is consistent with that shown in \cite{stacy&bromm2013}, and is also similar to that of previous work (e.g. \citealt{yoshidaetal2006,greifetal2011}).  
The minihalo forms at $z\sim28$ with a mass and virial radius of $3 \times 10^{5}$ M$_{\odot}$ and 70 pc.  The gas heats through adiabatic compression as it approaches densities of 10$^8$ cm$^{-3}$.  After this point, three-body reactions rapidly increase the H$_2$ fraction, and the increased cooling rate is similar to the adiabatic heating rate due to collapse, yielding a roughly isothermal evolution.  The gas becomes fully molecular once it reaches densities of 10$^{10}$ cm$^{-3}$, after which the H$_2$ cooling is no longer optically thin and the gas gradually heats again to $\sim$\,2000 K as it collapses to $n = 10^{16}$ cm$^{-3}$.  The first sink particle forms
within a $\sim$\,1\,M$_{\odot}$ molecular core.  

Fig. \ref{velprof} shows the velocity structure of the high-density gas immediately prior to the formation of the first protostar.  The inner 1000 AU is dominated by radial infall and has a radial velocity $v_{\rm rad}$ that is greater than the rotational velocity  $v_{\rm rot}$.  The gas between 1000 and 10$^5$ AU is more rotationally supported with $v_{\rm rot} > v_{\rm rad}$.  The gas beyond 1000 AU is also marked by a higher turbulent Mach number $M_{\rm turb}$, defined as

\begin{equation}
M_{\rm turb}^2 c_s^2 =\left\langle\left(\vec{\rm v} - \vec{\rm v}_{\rm rot }  - \vec{\rm v}_{\rm rad  }\right)^2\right\rangle \mbox{,}
\end{equation}

\noindent where $c_s$ is the sound speed, $\vec{\rm v}$ the total velocity, and where we perform a mass-weighted average over all SPH particles in a radial bin (\citealt{stacy&bromm2013}). This transition region between 1000 and 10$^4$ AU marks where a secondary clump of gas later collapses out of an elongated structure and forms stars.  This occurs $\sim$\,1000 yr after the initial gas clump has already formed several sinks (see discussion in following sections).  
This secondary density peak is also apparent in the upper-left panel of Fig.~\ref{toomre} (we discuss this figure in greater detail below).  

The secondary clump formation is similar to that described in \cite{turketal2009}.  As in their work, we find that at $\sim 1000$ AU the ratio of rotational to gravitational binding energy becomes greater than 0.44, leading to rotational instability and the eventual gas collapse in this region (upper left panel of Fig. \ref{toomre}).  
At the point when the primary sink forms, the gas density in the secondary clump is between 10$^9$ and 10$^{10}$ cm$^{-3}$ and has a free-fall time of $\sim 1000$ yr -- consistent with time it takes for the secondary clump to collapse.

We may compare the properties of our simulated gas cloud to predictions of semi-analytic models (e.g. \citealt{tan&mckee2004}).
We measured $f_{\rm Kep} = v_{\rm rot}/v_{\rm Kep}$ and the sound speed $c_s$ over a wide range of enclosed mass $M_{\rm enc}$, where $v_{\rm Kep}$ is the Keplerian velocity.  
We find that between $M_{\rm enc} =$ 10 and 1000 M$_{\odot}$, $f_{\rm Kep} \sim 0.5$ while ranging from 0.4 to 0.8.  
We additionally estimate the entropy parameter $K'$, defined in \cite{tan&mckee2004} as

\begin{equation}
K' = \frac{P/ \rho^{\gamma}}{1.88 \times 10^{12} {\rm cgs}} = \left(\frac{T}{300 {\rm K} } \right) \left( \frac{10^4 \rm cm^{-3}} {n_{\rm H}}\right)^{0.1} \mbox{.}
\end{equation}

\noindent Not accounting for rotation or feedback, this can be used to evaluate their analytical estimate of the accretion rate onto the star-disc system:

\begin{equation}
\dot{M} = 3.2 \times 10^{-3} {\rm M_{\odot}\,yr^{-1}} \frac{K'^{15/7}}{\left( M_*/100 {\rm \,M_{\odot}}\right)^{3/7}} \mbox{\, .}
\end{equation}

\noindent For our case we estimate $K' \sim 1.2$.  This corresponds to accretion rates ranging from  $3 \times 10^{-2}$ to $\sim 10^{-2}$ M$_{\odot}$ yr$^{-1}$ for $M_* = 1$ M$_{\odot}$ to  $M_* = 10$ M$_{\odot}$.  
This is consistent with accretion onto the protostar found in the simulation at early times, but the analytic estimate overestimates the measured accretion rate at later times when secondary protostars have formed and feedback effects become stronger.  
From \cite{mckee&tan2008} equation (12), we may furthermore estimate that with no feedback this protostar will grow to $\sim$\,300\,M$_{\odot}$.  

 \cite{hiranoetal2014} accounted for the additional effects of feedback and angular momentum, and used their set of simulations to find a fit between primordial cloud mass $M_{\rm cloud}$, rotation parameter $\beta_{\rm cloud}$, and final stellar mass $M_{\rm fin}$.
$M_{\rm cloud}$ is the mass enclosed within the cloud radius $R_{\rm cloud}$, the distance where the ratio of $M_{\rm enc}$ to $M_{\rm BE}$ reaches a maximum.  
$M_{\rm BE}$ is the Bonnor-Ebert mass, which we measure just prior to the formation of the initial sink:

\begin{equation}
M_{\rm BE} \sim 1000 {\rm \,M_{\odot}} \left(\frac{T}{200 \rm K}\right)^{3/2}   \left(\frac{n_{\rm H}}{10^4 \rm cm^{-3}}\right)^{-1/2} \mbox{.}
\end{equation}

\noindent In our case $M_{\rm cloud} \simeq$ 600 M$_{\odot}$.  We apply this to the relation presented in \cite{hiranoetal2014}:

\begin{equation}
M_{\rm fin} = 100 {\rm \,M_{\odot}} \left(\frac{M_{\rm cloud}}{350 \rm M_{\odot}} \frac{0.3}{\beta_{\rm cloud}}\right)^{0.8} \mbox{.}
\end{equation}

\noindent $\beta_{\rm cloud}$ is defined as

\begin{equation}
\beta_{\rm cloud} = \frac{\Omega_{\rm cloud}^2 R_{\rm cloud}^3} {3 G M_{\rm cloud}} \mbox{,}
\end{equation}

\noindent where $\Omega_{\rm r} = v_{\rm rot}(r)/r$ at radius $r$ and $\Omega_{\rm cloud}$ is taken at the cloud radius. 
For our case $\beta_{\rm cloud} \sim 0.24$, yielding $M_{\rm fin} \sim 200$\,M$_{\odot}$.  In later sections we will further compare the simulation results to these estimates.
\
\begin{figure*}
\includegraphics[width=.85\textwidth]{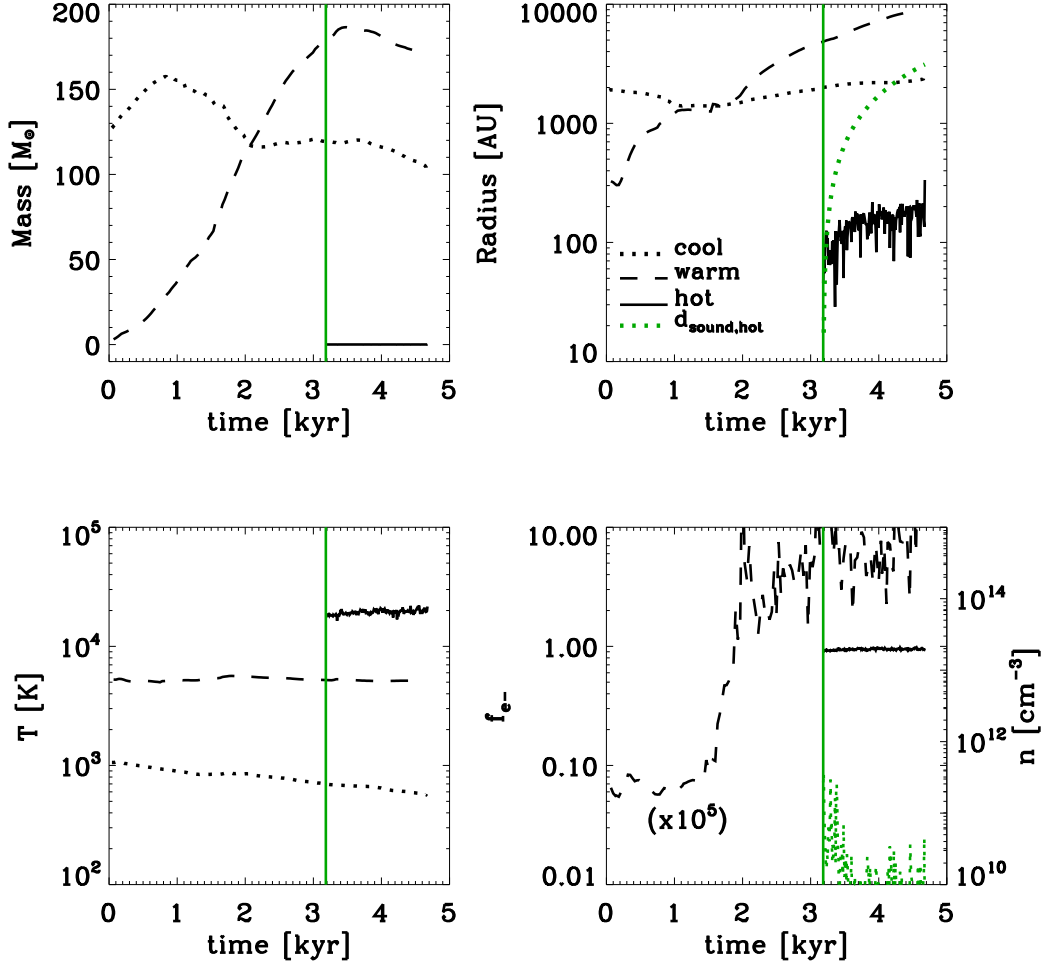}
 \caption{ 
Evolution of the disc gas phase (dotted lines), warm gas phase (dashed lines), and the I-front expanding from the main sink particle (solid black lines).  
The disc phase is defined as gas with $n > 10^9$  cm$^{-3}$ and $f_{\rm H_2} > 10^{-3}$.
The warm phase is defined as neutral gas with $T > 3000 $ K and $n > 10^{5}$ cm$^{-3}$.
The H{\sc ii} region first begins to grow at $t=3200$ yr, as marked by the solid blue vertical lines.  
{\it Upper Left:} Total mass of particles within the various phases.
{\it Upper Right:} Average distance of the various phases from the main sink.  The blue dotted line shows the hypothetical distance $d_{\rm sound, hot}$ traveled by a wave propagating at a sound speed of 10 km s$^{-1}$ from the time of I-front breakout.  In comparison, the I-front is confined and does not expand at this sound speed.
{\it Lower Left:} Average temperature of the gas phases.
{\it Lower Right:}  Average ionization fraction within the warm phase (dashed line), multiplied by a factor of $10^5$.
Solid line shows average ionization fraction within the  H{\sc ii} region, and green dotted line depicts number density. 
}
\label{if_evol}
\end{figure*}

\subsection{Disc Evolution and Fragmentation}

The evolution of the disc mass, radius, and temperature are shown in Fig. \ref{if_evol}.  For simplicity we define the disc as gas particles (sinks excluded) with $n > 10^9$  cm$^{-3}$ and $f_{\rm H_2} > 10^{-3}$.  This captures the cool molecular gas that has not been heated through protostellar feedback or radial infall onto the disc.  The disc mass $M_{\rm disc}$ grows steadily for the first 1000 yrs after initial sink formation, growing from initially 130\,M$_{\odot}$ to reach a peak of 160 M$_{\odot}$.  This averages to an infall rate onto the disc of $3 \times 10^{-2}$ M$_{\odot}$ yr$^{-1}$. However, this mass is lost rapidly to sink formation between 1000 and 2000 yr (see discussion in following section).  After this point gas flow into the disc balances out the accretion of disc gas onto the sinks, so that the total $M_{\rm disc}$ remains steadily at $\sim$\,120\,M$_{\odot}$ between 2000 and 4000\,yr.  At this transition point of 2000 yr, there is a rapid decline in accretion rate onto both the disc and the sinks.  This occurs as the warm phase of neutral $T\sim7000$ K gas expands at approximately the sound speed $c_s \sim 5$ km s$^{-1}$  (see figures and discussion in Section 5).  The warm phase reaches the disc radius of 2000 AU by 2000 yr, leading to a subsequent slowing of the accretion rate onto the disc and sinks.

\begin{figure*}
\includegraphics[width=.85\textwidth]{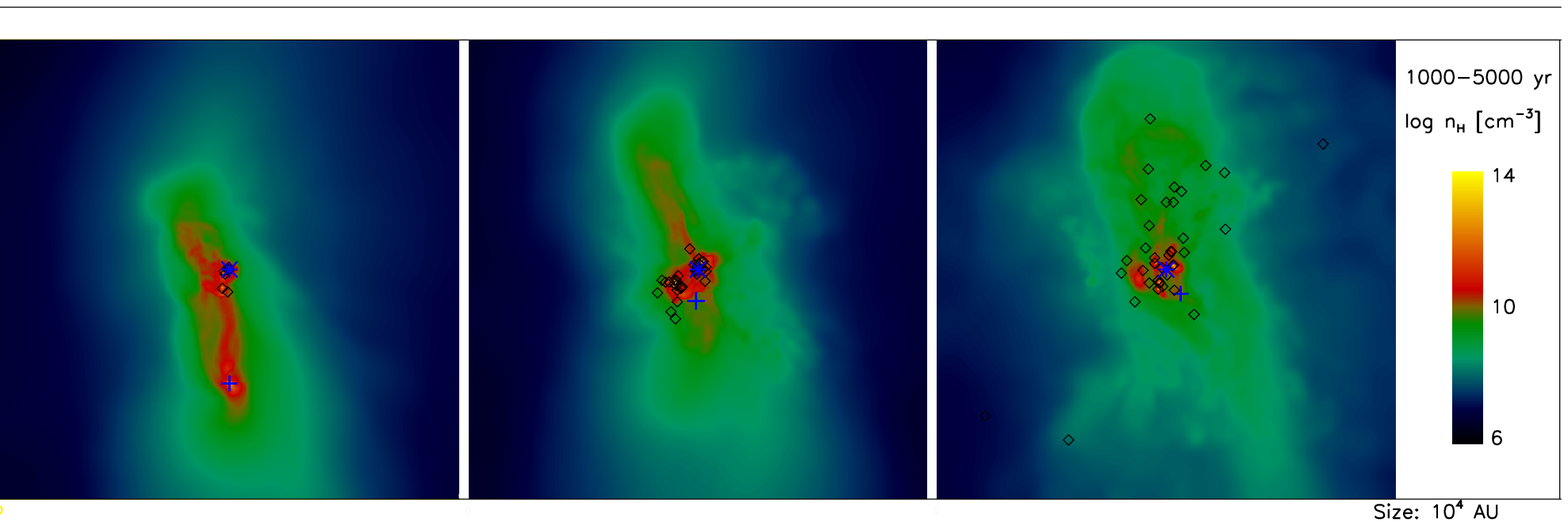}
\includegraphics[width=.85\textwidth]{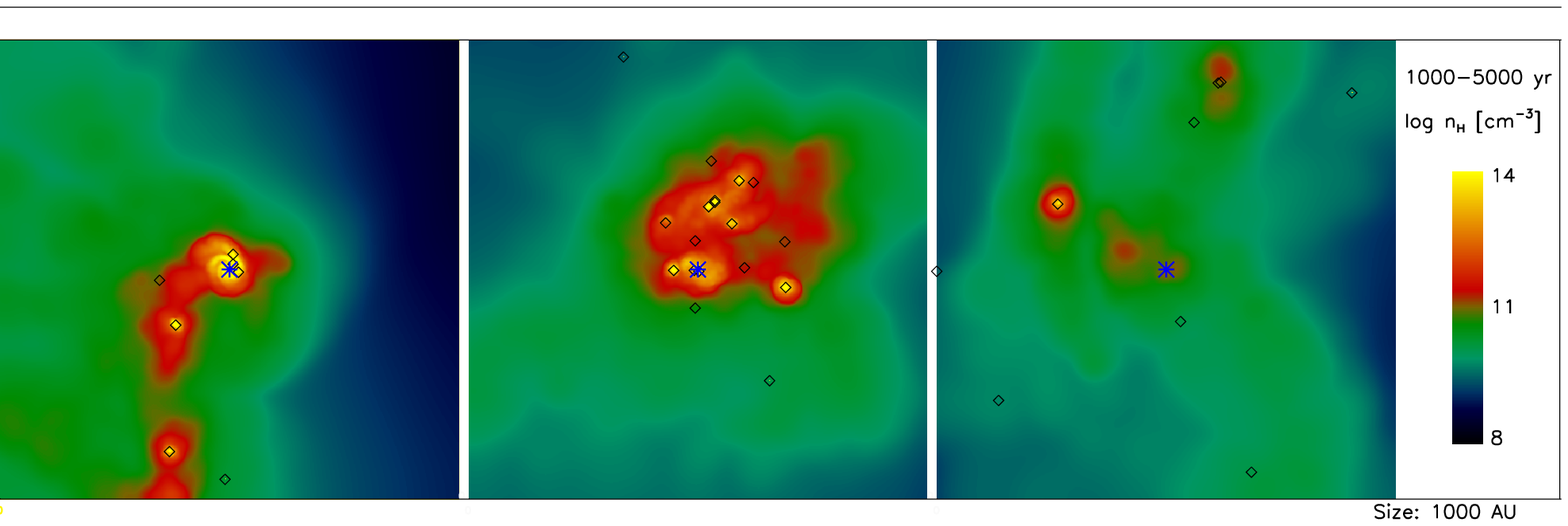}
\caption{
Density slice along the simulation x-y plane
of the central 10$^4$ AU (top row) and 1000 AU (bottom row).  
From left to right, structure is shown at 1000, 2000, and 5000 yr.
The early elongated structure gradually collapses into two distinct clumps by 1000 yr.  
The blue
asterisk marks the location of the largest sink.  
Black diamonds represent other secondary sinks.
The blue plus sign marks the first sink to form in the secondary clump, where its initial large separation from the other sinks is most visible in the top left panel.
These later merge into a larger disc structure after 2000 yr.
}
\label{nh-morph-1e4}
\end{figure*}

\begin{figure*}
\includegraphics[width=.8\textwidth]{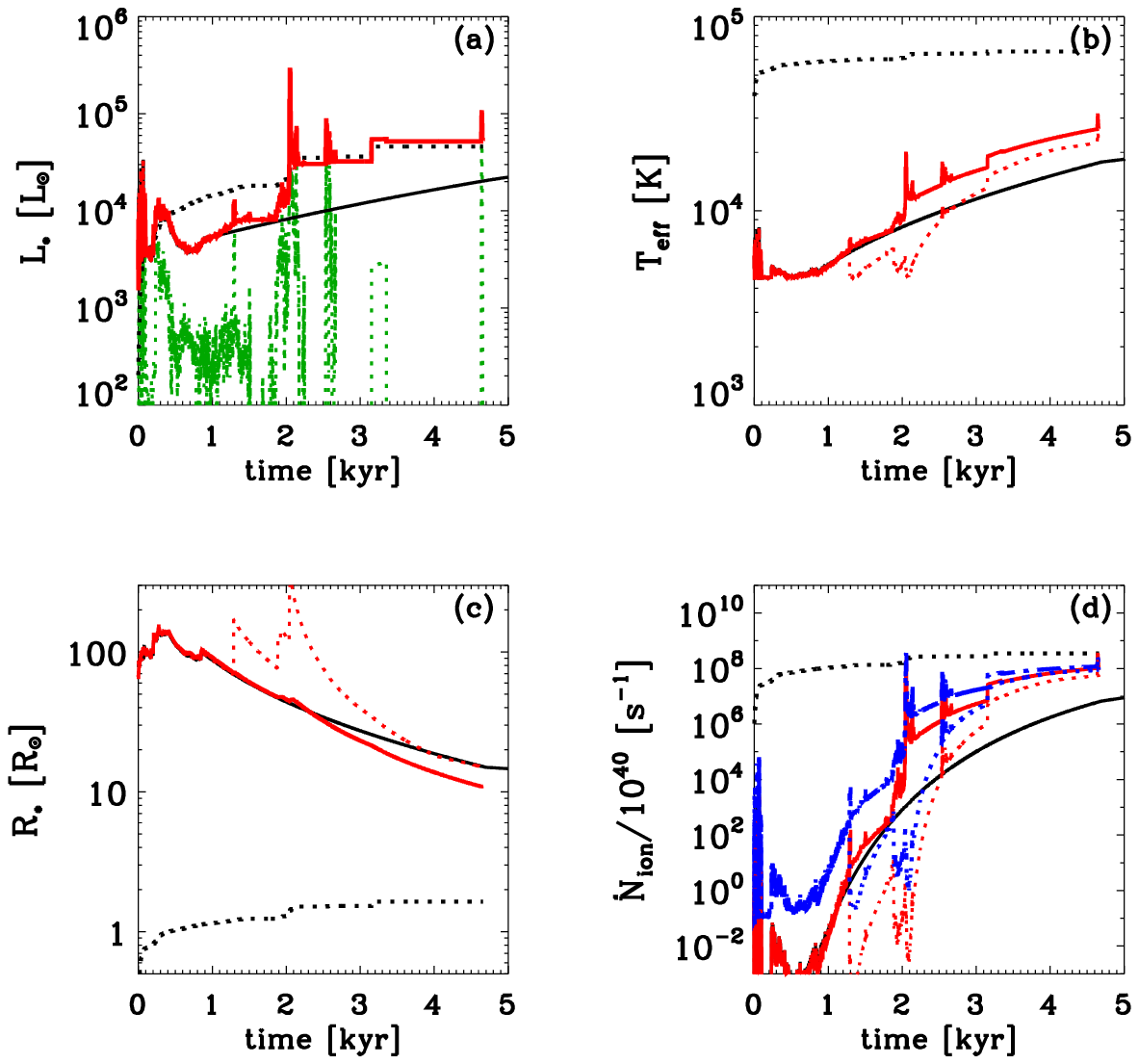}
 \caption{Evolution of various properties of the growing protostar according to our analytical model.  
Solid red lines represent the evolution of the protostar in our simulation, as determined by the sink mass and accretion rate.
Dotted red lines show the corresponding evolution when the limit on the protostellar expansion rate (Eq. 29) is not imposed.
Solid black lines represent the protostellar evolution when the model is switched to a constant $\dot{M} = 10^{-3}$ M$_{\odot}$ yr$^{-1}$ and $\alpha=1$ at 
$t_{\rm acc} = 1000$ yr instead of the variable accretion rate and $\alpha$ found from the simulation.
Black dotted lines represent the corresponding ZAMS stellar values for the current sink mass.  
{\it (a):} Total protostellar luminosity.   
Green dotted line is the accretion luminosity.
{\it (b):} Effective temperature.  
{\it (c):} Protostellar radius.
{\it (d):} Ionizing luminosity, $\dot{N}_{\rm ion}$.  
Solid blue line is the LW radiation luminosity in our model, while dotted blue line is the LW luminosity when we do not apply the limit on protostellar expansion.
}
\label{star-model}
\end{figure*}

As evident in Fig. \ref{nh-morph-1e4}, the disc gas does not actually maintain an axisymmetric structure, and even has a secondary dense clump which forms over 1000 AU from the main density peak where the first sink forms.  We thus simply define the disc radius as the average distance of all disc gas from the centre-of-mass of the sinks.  Throughout the simulation the disc radius remains at roughly 2000 AU, while the average gas temperature declines from 1000 to several hundred Kelvin. 

We employ the Toomre criterion to examine the instability of the primordial disc to fragmentation:

\begin{equation}
Q = \frac{c_{\rmn s} \kappa}{\pi G \Sigma} < 1  \mbox{\ .}
\end{equation}

\noindent Here, $\Sigma$ is the disc surface density and $\kappa$ the epicyclic frequency, which is equal to the angular velocity for Keplerian rotation 
$\Omega_{\rm Kep} = \sqrt{G M_{\rm enc}/r^3}$.
As described in \cite{stacyetal2014}, we estimate $\Sigma(r_i) \sim M_{\rm shell}(r_i) / [4 \pi (r_i^2 - r_{i-1}^2)]$.  $M_{\rm shell}(r_i)$ is the mass enclosed within a spherical shell with inner and outer surfaces spanning $r_{i-1}$ to $r_i$, oriented perpendicular to the disc rotational axis and centred on the main sink.  Warm particles with $T > 3000$ K are excluded from the mass total.   
At the time of initial sink formation, within the central several thousand AU the measured angular velocity is consistent with
 $\Omega_{\rm Kep}$ to within a factor of less than two.

Fig. \ref{toomre} shows 
the Toomre Q values
at various times in the simulation.  At $t=0$, the time of initial sink formation, the disc satisfies $Q<1$ at distances from  10 to over 1000 AU from the sink.  
The resulting formation of sinks and spiral arms is visible in 
the bottom-left panel of
Fig. \ref{nh-morph-1e4}.

As spiral structure develops at this time, the collapse of gas into a secondary sink first occurs at $t=400$ yr 
at a distance of $\sim$\,400 AU away from the first sink (see also Section 6) .  
As described in \cite{dopckeetal2013}, we can compare to the estimate for fragmentation timescale:

\begin{equation}
t_{\rm frag} \sim M_{\rm BE} / \dot{M}_{\rm sphere} \mbox{,}
\end{equation} 

\noindent where $\dot{M}_{\rm sphere}$ is defined as

\begin{equation}
\dot{M}_{\rm sphere} = 4 \pi r^2 \rho v_{\rm rad} \mbox{.}
\end{equation} 

\noindent The timescale over which the central star-forming cloud grows in mass can be approximated by

\begin{equation}
t_{\rm growth} \sim M_{\rm enc} / \dot{M}_{\rm sphere} \mbox{,}
\end{equation} 

\noindent and we can expect fragmentation to occur more rapidly than accretion where $t_{\rm frag} / t_{\rm growth} < 1$.  
Fig. \ref{toomre} shows estimates for $\dot{M}_{\rm sphere}$,  $t_{\rm frag}$, $t_{\rm growth}$, and $t_{\rm frag} / t_{\rm growth}$ at the time of initial sink formation.   
From Fig. \ref{toomre} we can see a dip in $t_{\rm frag} / t_{\rm growth}$ below unity at $r \ga 100$ AU, where $t_{\rm growth}$ is 
indeed of order a few hundred years.  This matches well the distance and time at which the second sink forms.  

The secondary density peak at $\ga$ 1000 AU from the main sink begins to form a cluster of sink particles at $\sim$\,1000 yr, coinciding with
the region where $Q < 1$ at this time. In Fig.~\ref{toomre}, we show how the region of Toomre instability propagates to progressively larger distances over the first 1000 yr, until it reaches the secondary clump.
The time at which the secondary clump begins to fragment is also consistent with our estimate of $t_{\rm frag}$ at $\sim$\,1000 AU to within a factor of two. 
As the disc evolves, less of the region around the main sink remains unstable to fragmentation (red line in Fig.~\ref{toomre}). 
By 3000 yr the rate of new sink formation has greatly declined (see Section 6).

\begin{figure*}
\includegraphics[width=.8\textwidth]{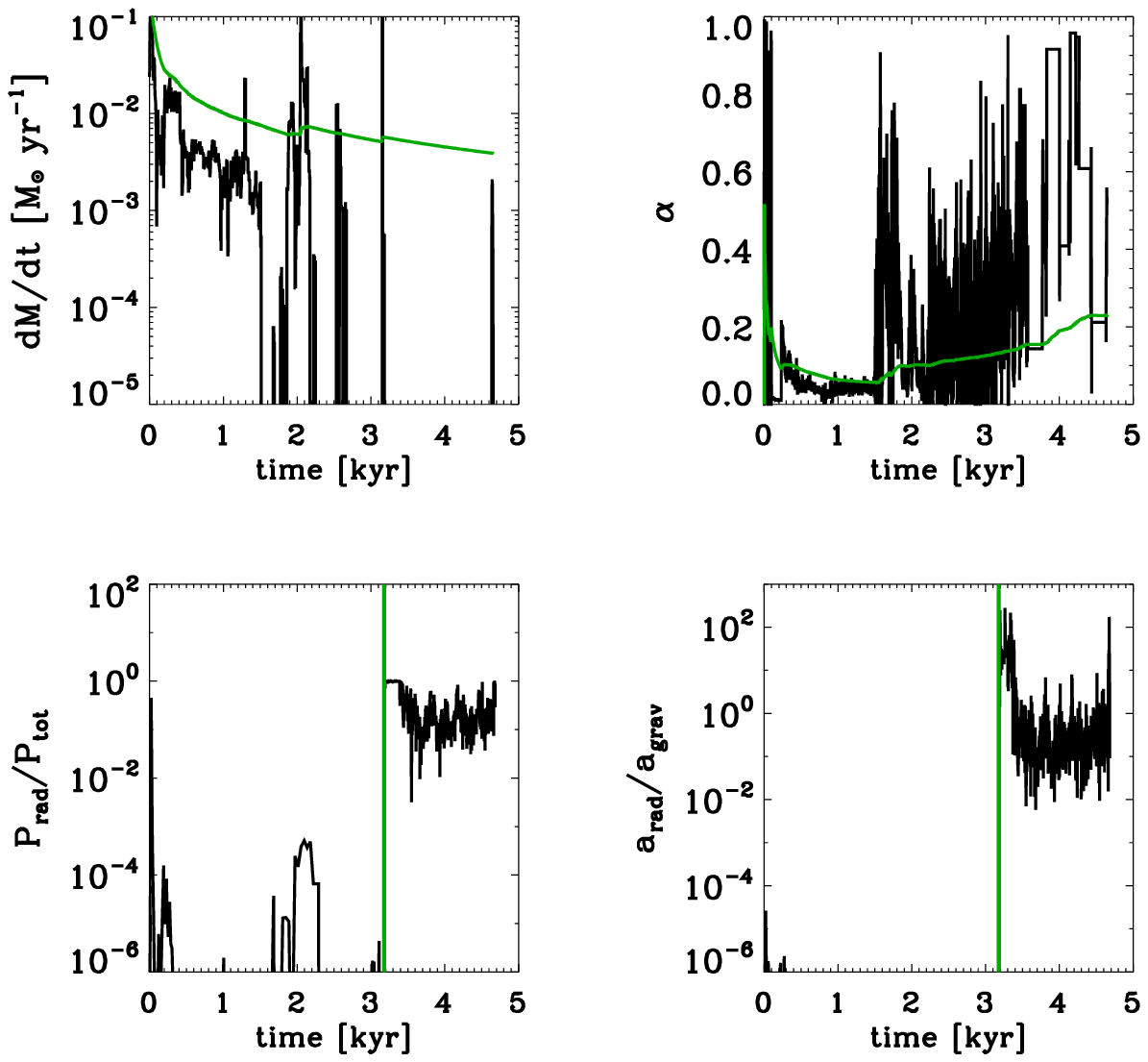}
 \caption{Evolution of various properties of near-sink gas:
 {\it (a):}  Accretion rate onto the main sink. 
{\it (b):} Value of $\alpha$, where $\alpha=1$ corresponds to gas flowing radially towards the sink, while $\alpha=0$ corresponds to a rotationally dominated flow.  
Green lines show the running linear 
averages over the simulation time.
{\it (c):} Ratio of ambient pressure from diffuse ionizing radiation $P_{\rm rad}$ to total the total pressure ($P_{\rm tot} = P_{\rm rad} + P_{\rm therm}$).  
Before break-out of the I-front, $P_{\rm rad}$ is averaged for all particles with $n > 10^{13}$ cm$^{-3}$.  After I-front break-out at $t=3200$ yr, denoted by the green vertical line, $P_{\rm rad}$ is averaged for all particles within the H{\sc ii} region.
{\it (d):}  Ratio of radially outward acceleration from direct ionizing radiation ($a_{\rm rad}$) to inward gravitational acceleration ($a_{\rm grav}$).  We apply $a_{\rm rad}$ only to particles within the H{\sc ii} region, while for all other gas $a_{\rm rad}=0$.
}
\label{star-model2}
\end{figure*}

\section{Onset of Radiative Feedback}

\subsection{Lyman-Werner Feedback}

Figs \ref{star-model} and \ref{star-model2} show the evolution of the protostar and nearby gas when our protostellar model is applied to the main sink.
At various points in the evolution, $L_*$ approaches or exceeds the $L_{\rm ZAMS}$ value for a star of the same given mass.  
However, $T_{\rm eff}$ remains well below $T_{\rm ZAMS}$ throughout the simulation since the stellar radius is still much larger than a main sequence star of the same mass.

At $\sim$\,2000 yr, the main protostar first reaches sufficiently large  $L_*$ and $T_{\rm eff}$ to cause an initial burst of radiative feedback.
Fig. \ref{if_evol} shows that this occurs when the mass and radius of gas in the warm phase first exceeds that in the cool disc phase. 
Because our protostellar model employed a limit to the rate of protostellar expansion (see Section 3.2.3), the radiative feedback grew significant at slightly earlier times.  
Fig. \ref{star-model} compares our model with one in which the protostellar expansion was not limited, and in this case $T_{\rm eff}$ reaches $2 \times 10^4$ K at $\sim$ 3000 yr instead of 2000 yr.  By $\sim$ 3500 yr, however, both models converge to similar values, implying that the overall long-term evolution of the disc will be similar in either case.

\begin{figure*}
\includegraphics[width=.8\textwidth]{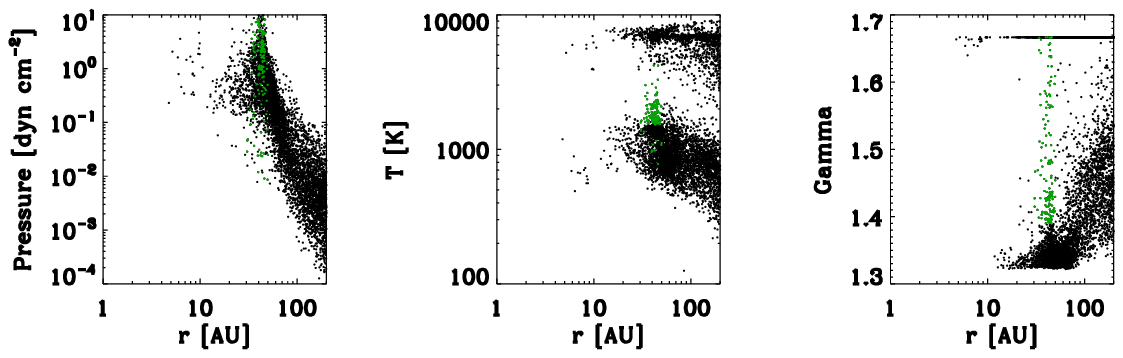}
\includegraphics[width=.8\textwidth]{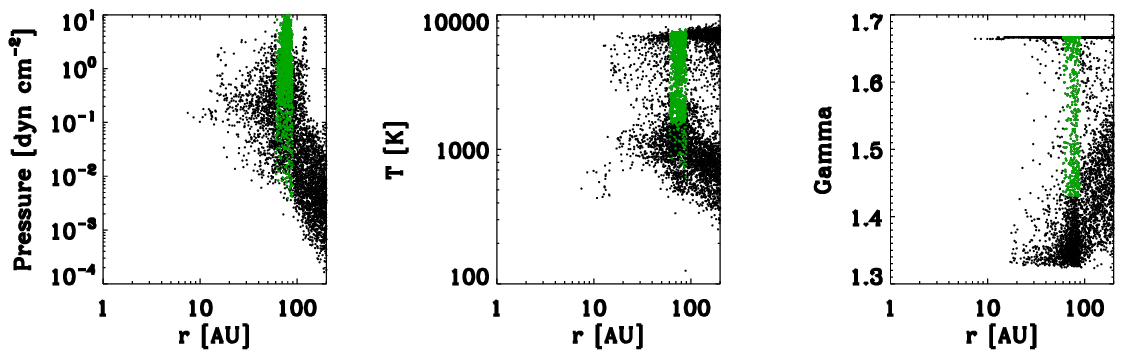}
 \caption{ 
Radial profile of gas properties around the sink at 1705 yrs (top row) and 1710 yrs (bottom row).
Particles which undergo rapid dissociation and change in adiabatic $\gamma$ due to LW feedback are marked in green.  These particles undergo a pressure wave which rapidly expands away from the sink.  
Note that these particles also quickly heat from $\sim$\,1000 to $\ga$\,5000 K.
}
\label{pwave}
\end{figure*}

For our given protostellar model, the onset of radiative feedback
coincides with a period of rapid accretion onto the main sink, triggered as the two separate clusters of sinks approach each other.  
As a result, the protostar's LW photon luminosity quickly grows from $\sim$\,10$^{44}$ to 10$^{48}$~s$^{-1}$ 
(Fig. \ref{star-model}).  
At this point the dissociating radiation causes a shell of particles around the sink to undergo a transition from molecular to atomic.  
Their corresponding adiabatic index $\gamma$ transitions from $\sim$\,7/5 to $\sim$\,5/3.
Along with molecular dissociation, this gas also undergoes heating from $\sim$\,1000 to $\ga$5000 K.
As discussed in \cite{susa2013}, the dissociation through LW radiation allows for H$_2$ formation heating to occur without the associated cooling through collisional dissociation, increasing the gas temperature.  
This combination of processes 
generates a pressure wave which travels outward from the sink (Fig. \ref{pwave}), corresponding to a boost
in pressure from about 1~dyn~cm$^{-2}$ to 10~dyn~cm$^{-2}$, initially occurring at a distance of 
$\sim$\,10~AU.

The smoothing length of the boosted particles is $h \sim 10^{14}$ cm, and their mass is 
$1.2 \times 10^{30}$ g.  
From this, we may estimate that the dissociated particles experience an acceleration boost:
\begin{eqnarray}
a_{\rm boost} &\sim& \frac {\Delta P \,  h^2}  {m_{\rm sph}}  \\
&\sim& \frac {\left(10 \, {\rm dyn \, cm^{-2}} \right)  \left(10^{14} {\rm cm} \right)^2 } {1.2\times 10^{30} {\rm g}} \sim 0.1 {\rm \,cm \, s^{-2}}\mbox{\,.}
\end{eqnarray}

\noindent In comparison, the gravitational acceleration is somewhat smaller:
\begin{equation}
a_{\rm gravity} \sim \frac {G M_*} { r^2} \sim  \frac {G \, 10 \rm M_{\odot} } { \left(10 {\rm AU}\right)^2} \sim 0.06 {\rm \,cm \, s^{-2} }\mbox{\,.}
\end{equation}

\noindent Over the 10 yr period during which this boost initially occurs, we thus find that
$a_{\rm total} \sim a_{\rm boost} - a_{\rm gravity} \sim 0.05 {\rm \, cm \, s^{-2}}$.  This accelerates the particles to a distance of roughly 
$d \sim \frac{1}{2} 0.05 \, {\rm cm \, s^{-2}}  \, \left(10 {\rm yr} \right)^2 \sim 2.5\times 10^{15} {\rm cm} \sim 160 \rm \, AU$.
This compares well with the simulation results (Fig. \ref{pwave}), as the pressure wave does indeed reach distances of $100 \, {\rm AU}$ within this time period.

\begin{figure*}
\includegraphics[width=.8\textwidth]{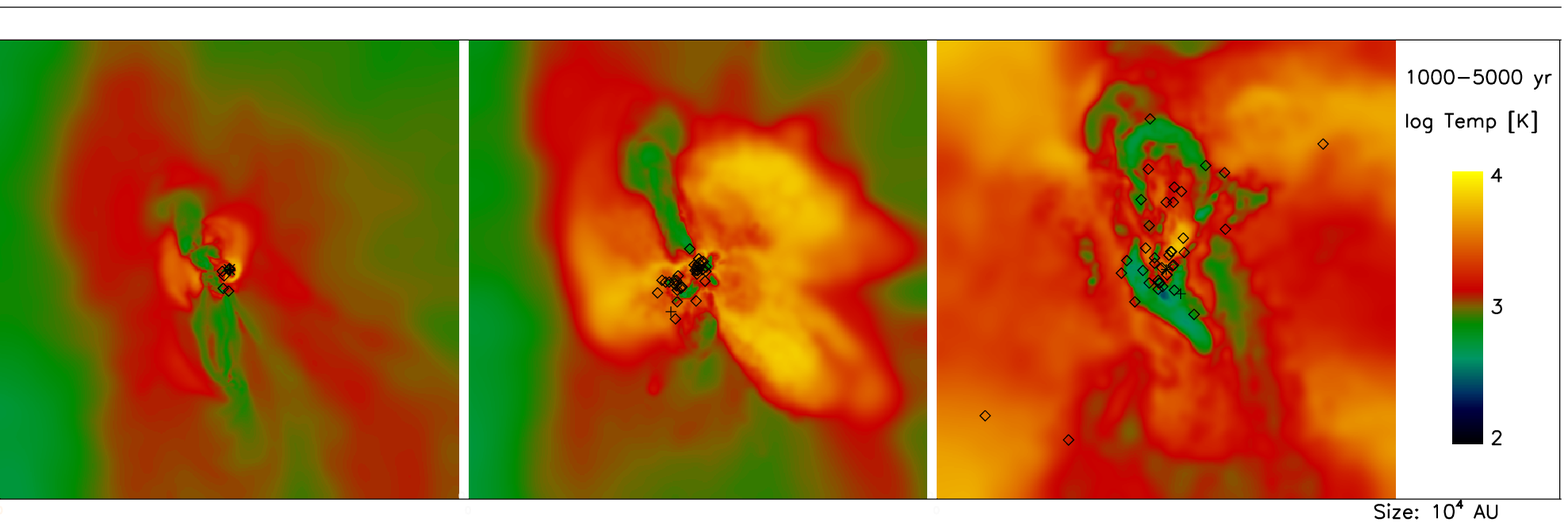}
\includegraphics[width=.8\textwidth]{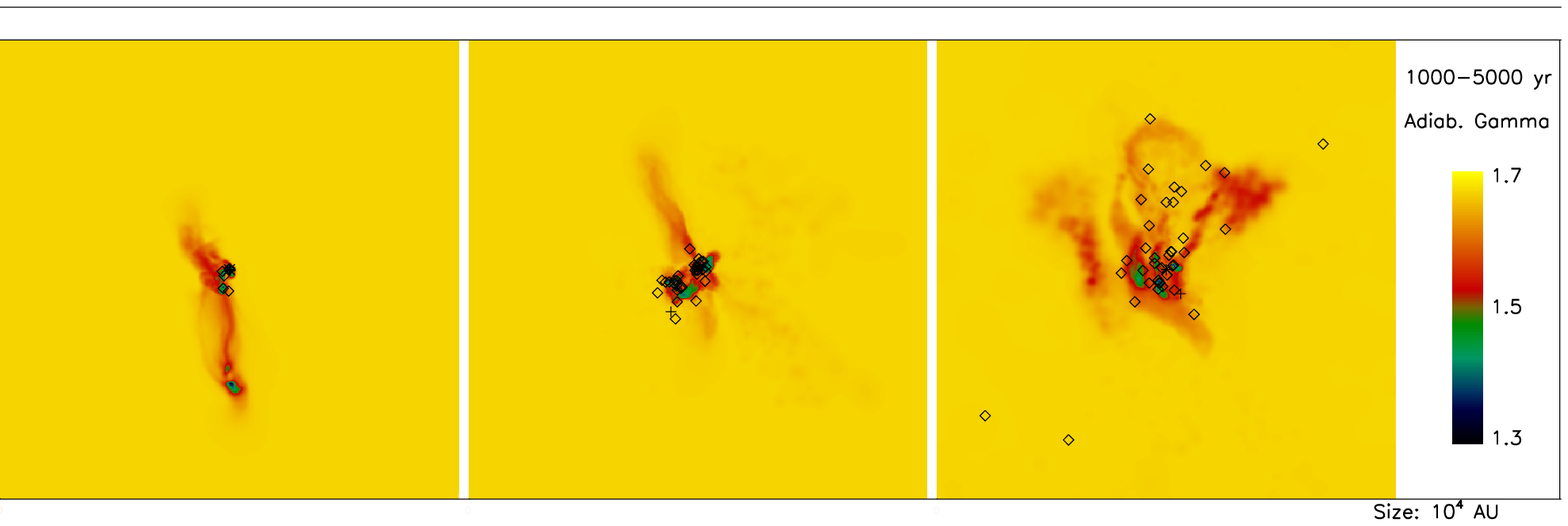}
 \caption{Same as top panel Fig. \ref{nh-morph-1e4}, but instead showing temperature in the top three panels and adiabatic $\gamma$ in the bottom three panels. 
 From left to right, structure is shown at 1000, 2000, and 5000 yr.
 }
\label{temp-morph-1e4}
\end{figure*}

This early acceleration easily increases the velocity of the pressure wave to $\la$ 100 km s$^{-1}$ within a few years.  In comparison, the escape velocity from the 15~M$_{\odot}$ sink ranges from $\sim$\,40 to 10 km s$^{-1}$ over distances of 10 to 100 AU.  The high velocity of the pressure wave thus allows it to coast unhindered by gravity, reaching distances of several thousand AU within a few hundred years (Fig. \ref{temp-morph-1e4}).  This greatly slows the mass growth of the disc and sinks even before the development of an H{\sc ii} region.

\subsection{I-front Breakout and Evolution}

\begin{figure}
\includegraphics[width=.45\textwidth]{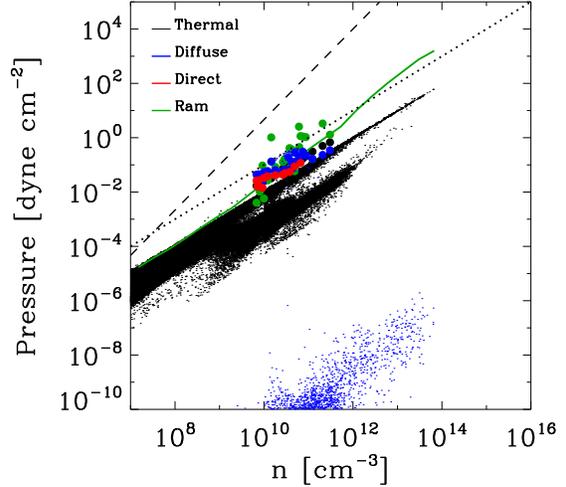}
 \caption{ 
 Pressure versus density at 3000 yr after the sink formation.
 Thermal pressure $P_{\rm therm}$ is denoted with black dots.  Diffuse and direct radiation pressure  ($P_{\rm rad,diff}$ and $P_{\rm rad,dir}$) are shown with blue and red dots, respectively.  
Dots are enlarged for ionized particles.
Average ram pressure $P_{\rm ram}$ is shown with the green line, while green dots show  $P_{\rm ram}$ for individual ionized particles.
For comparison, dashed and dotted black lines show $n^{5/3}$ and $n^1$ powerlaws, respectively.
Note how the gas follows a $P_{\rm therm} \propto n$ relation, but with a  higher normalization for hotter gas. 
Three separate gas phases - cool molecular, warm neutral, and hot ionized - are apparent in the thermal pressure.  
For the neutral gas, $P_{\rm rad,dir}$ is set to zero while $P_{\rm rad,diff}$ is negligible, as shown by the lower blue dots.
Within the densest ionized gas, both $P_{\rm rad,diff}$ and $P_{\rm rad,dir}$ are similar to $P_{\rm therm}$  (large upper blue, red and black dots).  
As the I-front expands to lower densities, radiation pressure will decline more rapidly than thermal pressure such that $P_{\rm therm}$ will again dominate at $n \la 10^7$ cm$^{-3}$.
}
\label{pressure}
\end{figure}

The I-front first appears 3200 years after the main sink forms.  Its subsequent evolution is shown in Fig. \ref{if_evol}.  From 3200 to 5000 yr the I-front expands from $\la$ 100 AU to $\sim$\,200 AU, while the mass of the ionized region fluctuates around 10$^{-2}$ M$_{\odot}$.  
The thermal, radiation, and ram pressures at 4000 yr are compared in Fig. \ref{pressure}.  
Here, we estimate ram pressure as 
$P_{\rm ram} \sim (1/2) \rho v^2$, 
where $v$ is the velocity relative to the main sink.  $P_{\rm ram}$ is greater than $P_{\rm therm}$ and similar to both $P_{\rm rad,diff}$ and $P_{\rm rad,dir}$.  
At this time the gravitational radius, the radius at which the sound speed of the ionized gas exceeds the free-fall velocity, is approximately

\begin{equation}
r_{\rm g} \sim \frac{G M_*} {c_s^2} \sim 200 {\rm  AU} \mbox{,}
\end{equation}

\noindent where $M_* \sim 20$ M$_{\odot}$ and $c_{\rm s} \sim 10$ km s$^{-1}$.  Therefore, the ionization front radius, $r_{\rm HII}$, is confined to  $r_{\rm HII} \la r_{\rm g}$, resulting in an ultra-compact configuration.
Due to a combination of ram pressure and gravitational confinement,
along with low particle resolution in comparison with the typical Stromgren radius (see section 2.4),
we find that the feedback from ionizing radiation does not extend to large distances, thus not affecting the evolution of the central gas or emerging Pop~III cluster within the first 5000~yr. 
However, as the main star continues to contract, both $\dot{N}_{\rm ion}$ and $r_{\rm HII}$ are expected to increase, further inhibiting stellar accretion.

\begin{figure*}
\includegraphics[width=.8\textwidth]{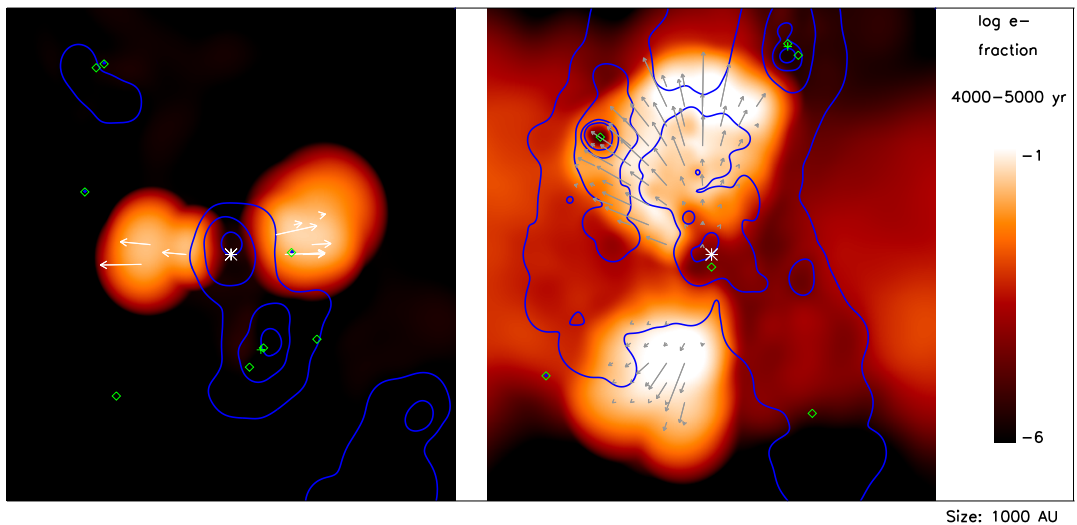}
\caption{
Ionization fraction of central 1000 AU of gas around first sink at 4000, and 5000 yr.  
Asterisks denote the location of the main sink.  Green diamonds are secondary sinks.
Blue lines denote density contours at $n = 10^{10}$, $3\times10^{10}$, $10^{11}$, and $3 \times 10^{11}$ cm$^{-3}$.
Arrows in left and right-most panels indicate the direction and magnitude of $a_{\rm ion,dir}$.
The confined H{\sc ii} region fluctuates in size and orientation as the density structure around the protostar evolves.
 }
\label{hii-morph}
\end{figure*}

Figure \ref{hii-morph} additionally shows the structure of the H{\sc ii} region, where it is apparent that the orientation of the H{\sc ii} region evolves rapidly.  This is due to the quickly changing configuration of gas and secondary sinks near the main protostar, which in turn alters the direction of greatest shielding.  
Rapid gas fragmentation can thus significantly alter the early development of the H{\sc ii} region.
However, at later times beyond 5000 yr when the i-front grows larger and the opening angle expands, the H{\sc ii} region may develop a preferred direction set by the density structure of the larger-scale envelope.

It will be beneficial for future work to more fully resolve the Stromgren radius at these high densities, to follow the i-front expansion for a longer period, and to trace the feedback from all sink particles.  This will provide more detail as to how significantly the ionized outflow alters the protostellar accretion beyond the already significant effects of photodissociation feedback.

\section{Emergence of the Pop III IMF}

\subsection{Protostellar Accretion Rate}

\begin{figure}
\includegraphics[width=.45\textwidth]{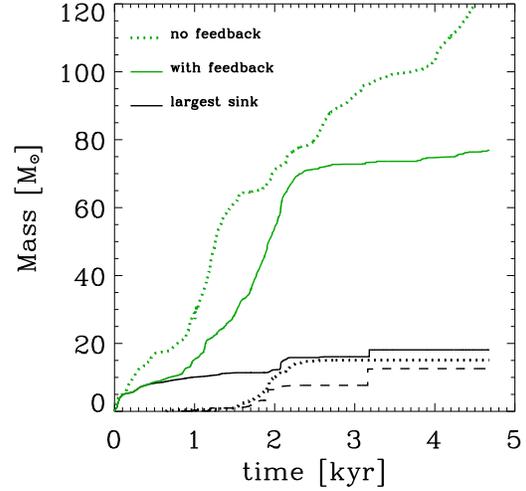}
 \caption{ 
 Sink growth over time.  Solid line represents the main sink.   Dotted and dashed lines show the sinks which reach the second and third-largest masses by the end of the simulation.  Solid green line is the total mass of all sinks.
Dotted green line is the total sink mass when simulated at lower resolution and without feedback (Stacy \& Bromm 2013).
 }
\label{sinkmass}
\end{figure}

The evolution of the total sink mass $M_{\rm sinks}$ and the growth of the three most massive sinks are shown in Fig. \ref{sinkmass}.  Over the first 1000 yr, $M_{\rm sinks}$ grows to $\sim$\,10 M$_{\odot}$, so $\dot{M}_{\rm sinks}$ is roughly $10^{-2}$ M$_{\odot}$ yr$^{-1}$. Between 1000 and 2000 yr, $\dot{M}_{\rm sinks}$ increases to $6\times10^{-2}$ M$_{\odot}$ yr$^{-1}$, accompanied by a burst of secondary sink formation.  
These early accretion rates are consistent with the analytical estimates from Section 4.1.

Between 2000 and 3000 yr, $\dot{M}_{\rm sinks}$ drops by a factor of ten to $6\times10^{-3}$ M$_{\odot}$ yr$^{-1}$, coinciding with the leveling off of $M_{\rm disc}$.
This flattening of the sink growth is not seen when simulating the same initial conditions but with lower resolution and no feedback (\citealt{stacy&bromm2013}), as depicted by the dotted red line in Fig. \ref{sinkmass}.  This further emphasizes the strengthening effect of feedback at 2000 yr.

At $t \sim 3200$ yr, the 16 M$_{\odot}$ main sink undergoes a significant merger event with a 2 M$_{\odot}$ secondary sink as their relative distance falls below 1 AU. We note that our technique of smoothing the accretion rate over 10 yr (see Section 3.1) may not realistically describe
the evolution of this stellar system, given our level of resolution. Merging may have instead occurred over an extended time of the order of $t_{\rm KH} \sim$ a few thousand years, possibly accompanied by a period of common envelope (CE) evolution (see, e.g.,  \citealt{pols&marinus1994, eggleton1983, hurleyetal2002}).
Given this uncertainty, we take the rate of mass flow between the stars to be 
$\dot{M_*} = 10^{-2}$ M$_{\odot}$ yr$^{-1}$, and we apply this rate to the protostellar evolution model for the following 200 yr.  This is equivalent to the primary accreting a total of 2 M$_{\odot}$, the mass of the secondary.  

In the latter part of the simulation after this merger event ($t > 3000$ yr), $M_{\rm sinks}$ gains just a few solar masses to grow to $\sim$\,80\,M$_{\odot}$, corresponding to $\dot{M}_{\rm sinks}$ of a few times $10^{-3}$ M$_{\odot}$ yr$^{-1}$.  
As this rate is expected to continue its decline due to I-front expansion, over the subsequent 10$^4$ to 10$^5$ yr the stellar system may only gain on the order of 10 M$_{\odot}$. Thus, the majority of accretion has already been completed by the end of our simulation, and the total sink mass may reach $M_{\rm fin} \sim$ 100 M$_{\odot}$. This is smaller but similar in magnitude to the estimate of $M_{\rm fin}$ predicted by \cite{hiranoetal2014}, though in contrast we find the total stellar mass is distributed amongst several tens of stars
instead of a single star.

\subsection{Protostellar Mergers and Ejections}

\begin{figure*}
\includegraphics[width=.45\textwidth]{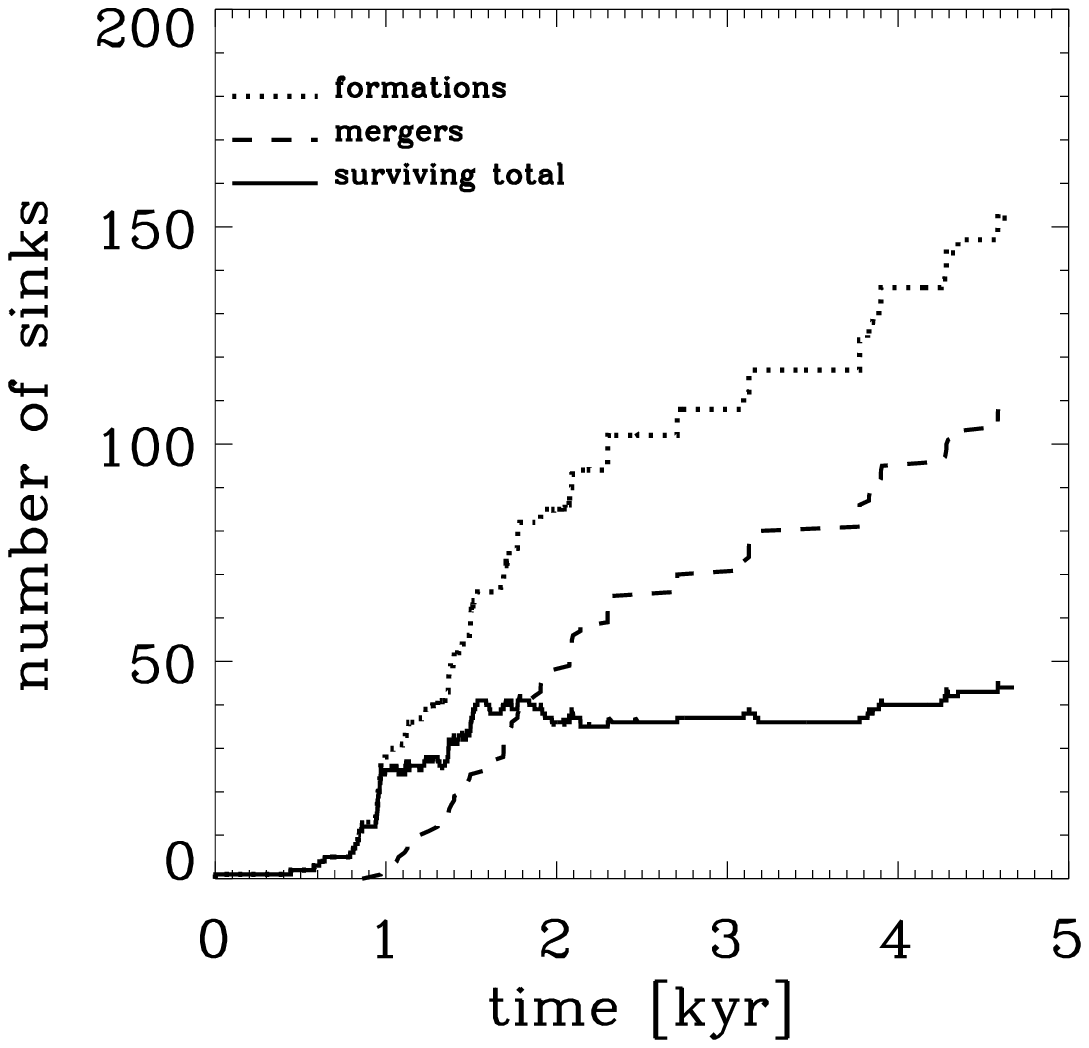}
\includegraphics[width=.45\textwidth]{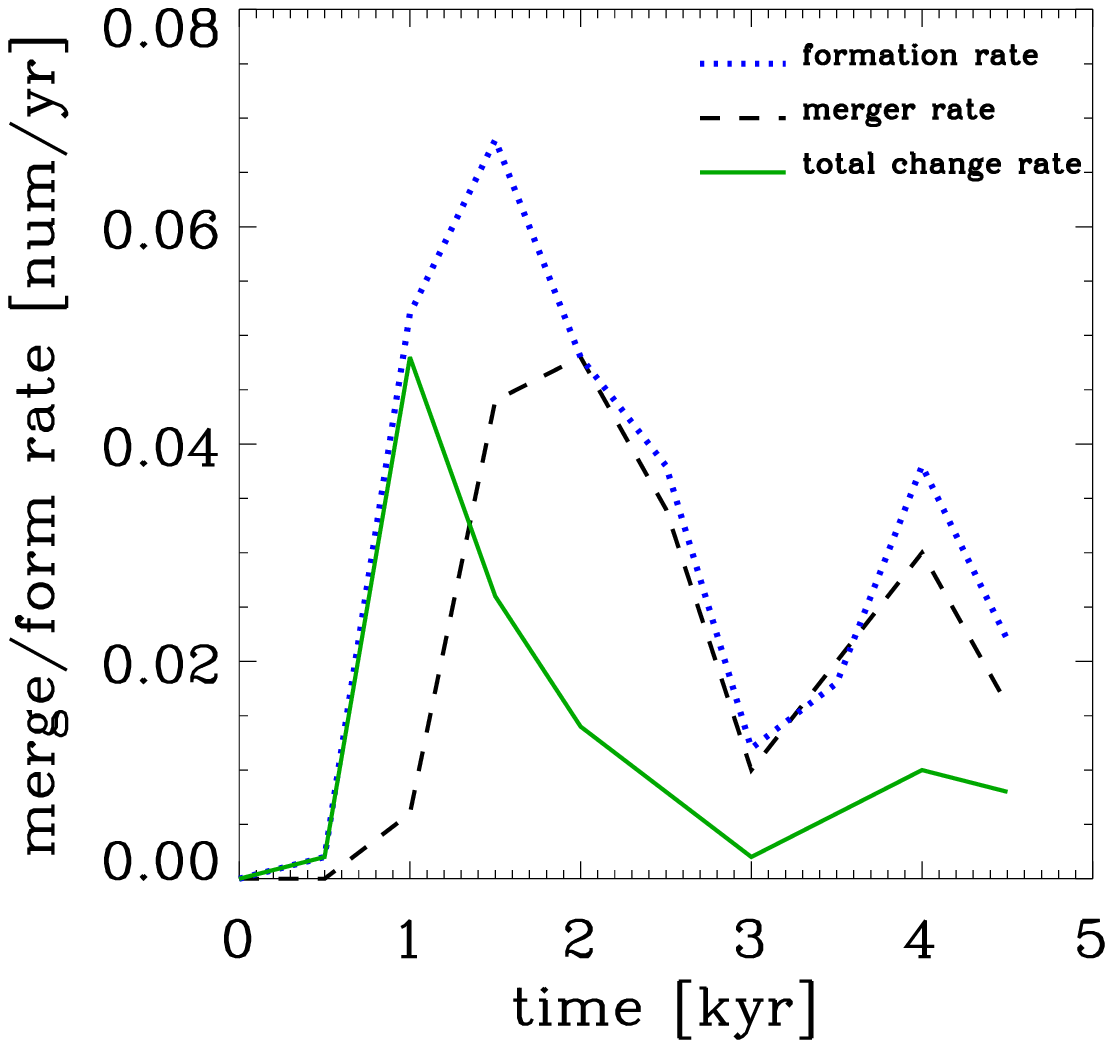}
 \caption{ 
{\it Left:} Cumulative sink number and mergers versus time.
Solid line depicts current number of surviving sinks over time.  Drops in sink number are due to sink mergers.  Dashed line shows total number of sink mergers over time.  The dotted line is the total number of sinks, both merged and surviving, that have formed in the simulation.
{\it Right:} Rate of formation of new sinks over time (blue dotted line) and rate of sink mergers (dashed black line).  Green line depicts rate of change in the  number of currently existing sinks.  As new sink formation and sink mergers occur at very similar rates after $\sim$\,2000 yr, the rate of change in total current sinks declines.  
}
\label{sinknum}
\end{figure*}

Secondary sink formation commences at 400 yr after the first sink has formed, while the majority of sinks forms between 1000 and 2000 yr.
The number of surviving sinks, $n_{\rm surv}$, refers to the number of sinks that currently exist in the simulation, thus excluding sinks that have been lost to merger events.  
The value of $n_{\rm surv}$ quickly grows to $\sim$\,40 within the first 1500 yr.
This corresponds to a growth rate of $\dot{n}_{\rm surv} \sim3 \times 10^{-2}$ yr$^{-1}$ (Fig. \ref{sinknum}).  
The lower resolution simulations of \cite{stacy&bromm2013} employed an accretion radius of $\sim$\,20 AU, while sink accretion was followed for 5000 yr in ten different minihaloes. There, the number of sinks formed ranged from four to 24. We note that the minihalo which formed 24 fragments was the lower-resolution version of the minihalo studied in this work.
In comparison, the investigation of \cite{greifetal2011} employed sink particles of a similar accretion radius ($\la$ 1 AU). Considering five different minihaloes, they found that 5 to 15 protostars formed over the 1000~yr followed in their simulations. \nocite{clarketal2011b} Clark et al. (2011b), also using $\sim$\,1~AU resolution, found an even higher fragmentation rate, with four protostars forming within only 100~yr.  
Hartwig et al. (2015b) examine four minihaloes with resolution of $\la$ 10 AU to find two to six sinks emerge in less than 1000 yr.

In contrast, the cosmological simulations of \cite{susaetal2014}, which employed a sink radius of 30 AU, found a lower number of between one and six sinks per minihalo
after longer times of $\sim$ 10$^5$ yr. 
More recently, \cite{hosokawaetal2016} observed disc fragmentation but no surviving secondary sinks within the five minihaloes studied at 30~AU resolution. In their work, any fragments instead rapidly migrate onto the central sink before they can further condense. We note that due to their limited resolution, any secondary sink formation within 30~AU of the central sink would be missed.
In summary, the rate of sink formation in our minihalo is thus rapid, but similar to that found in the previous highest-resolution simulations.  
Differences in formation rate likely stem from difference in resolution scale along with statistical variation between minihaloes.  Given that other simulations of similar resolution also find greater numbers of Pop III stars forming per minihalo, variation in numerical techniques may be a factor as well, and this should be examined in future work.

Our sink merger rate is also high, with roughly 70 mergers occurring within the first 2000~yr. The simulation of \cite{greifetal2012} explored primordial star formation with an extremely high spatial resolution of 0.05 R$_{\odot}$. Over the 10 yr simulated, approximately one out of three protostars survived to the end of the simulation. Interestingly, we find a similar trend over the first 2000~yr, where a total of 102 sinks formed, but only 37 remained while the rest were lost to mergers -- a survival rate of $\sim$\,36\%.    
This is high compared to \cite{hosokawaetal2016}, who at 30~AU resolution found a zero survival rate.  
Our high resolution allows us to resolve close encounters in which a fraction of the sinks survive.

After $t=2000$~yr, the value of $n_{\rm surv}$ remains roughly constant at $\sim$\,40, with new sinks formed balancing out sinks lost to mergers (Fig. \ref{sinknum}).  
Both the sink formation and merger rate decline after 2000 yr.  We estimate $\dot{n}_{\rm form}$ to be 5$\times$10$^{-2}$ yr$^{-1}$ and $\dot{n}_{\rm merge}$ to be 3$\times$10$^{-2}$ yr$^{-1}$ over the initial 2000 yr.
Afterwards, these rates both drop to $\dot{n}_{\rm form}$ $\sim$ $\dot{n}_{\rm merge}$ $\sim$ 2$\times$10$^{-2}$ yr$^{-1}$. 
This coincides with the time when the warm neutral medium reaches the disc radius of 2000~AU and the disc mass stabilizes at 120 M$_{\odot}$, as well as when the total sink mass becomes more steady at $\sim$\,75\,M$_{\odot}$.
Secondary sinks experience a variety of orbits before merging with a larger sink,
exhibiting orbital radii that range from $\ga$ 1 to $\sim$\,1000~AU prior to the merger event.  
In addition, some secondary sinks initially form at distances $< 30$ AU from other sinks, demonstrating the small resolution scales necessary to resolve all protostar formation and mergers.

We may estimate the expected merger timescale by considering the time over which pressure and gravitational torques operate, following \cite{greifetal2012}. Using the gravitational and hydrodynamic accelerations, $\bm{a}_{\rm grav}$ and $\bm{a}_{\rm hydro}$, we calculate the respective torques to find that gravity dominates over pressure on most scales. Specifically, torques are calculated as, e.g.,

\begin{equation}
 \bm{\tau}_{\rm grav} = \bm{r}_i \times \left( m_i  \bm{a}_{\rm grav} \right)\mbox{\ ,}
\end{equation}

\noindent where $m_i$ is the mass of the SPH or secondary sink particle and $\bm{r}_i$ is its distance from the main sink.  We then compare the dominating gravitational torque $\bm{\tau}_{\rm grav}$ to the angular momentum of the particle, 

\begin{equation}
\bm{l}_i = \bm{r}_i \times \left(m_i \bm{v}_i \right) \mbox{.}
\end{equation}

\noindent The corresponding timescale on which  $\bm{\tau}_{\rm grav}$ acts is then

\begin{align}
t_{\rm grav} = \frac{| \bm{l} | ^2}{\bm{l} \cdot \bm{\tau}_{\rm grav}}\mbox{\ .} 
\end{align}

\noindent For both SPH and sink particles $t_{\rm grav}$ ranges from roughly 10 to 1000 yr.

Consistent with the orbital histories observed in our simulation, sinks do indeed form and subsequently merge over timescales of hundreds to $\sim$\,1000 yr.  
In comparison, the orbital timescale is

\begin{equation}
t_{\rm orb} = \sqrt{ \frac{d^3} {G M_* } } \mbox{.}
\end{equation}

\noindent For $M_* = 10$ M$_{\odot}$ and $d$ ranging from 10 to 1000 AU, $t_{\rm orb}$ ranges from 1 to 1000 yr.  Sinks may undergo several orbits before merging, while interactions with gas and other sinks can increase instead of decrease angular momentum.  Thus, mergers generally occur over one to a few times $t_{\rm orb}$.

In contrast to mergers, close sink interactions can also lead to ejections from the stellar cluster.  
Similar to the analysis in \cite{stacy&bromm2013}, we examine which sinks may escape the disc and halo through N-body interactions.
Of the total of 45 sinks remaining at the end of our simulation, 14 have radial velocities that are greater than their escape velocities, where we define $v_{\rm esc}$ as

\begin{equation}
v_{\rm esc} = \sqrt{\frac{2 G M_{\rm enc}}{r}} \mbox{.}
\end{equation}

\noindent In this case $M_{\rm enc}$ is the mass enclosed between the disc centre of mass (COM) and the sink particle at distance $r$ from the 
COM.  The COM is determined using all gas and sink particles with $n > 10^{12}$ cm$^{-3}$.
Roughly  31\% of the sinks can thus leave the stellar disc. This is similar to the rate found in, e.g., \cite{greifetal2011} and \cite{stacy&bromm2013} in their simulations of primordial star formation, and \cite{bateetal2003} in their numerical studies of present-day, low-mass star formation.  
Of these 14 escaping sinks, 11 have 
$M_* < 1$ M$_{\odot}$, thus having the potential to survive to the present day.
In addition, we note that the velocity required to escape the minihalo altogether is $\sim$\,5 km s$^{-1}$, which is smaller than the disc escape velocity, so sinks that escape the disc can escape the minihalo as well.

\subsection{Binary Statistics}

A number of the sinks are in binary pairs at the end of the simulation. We establish binarity in the same way as described in \cite{stacy&bromm2013}, looking for pairs of sinks which are bound and have negative orbital energy.  For each pair we determine their specific orbital energy and kinetic energy, $\epsilon_{\rm p}$ and $\epsilon_{\rm k}$.   Binaries are defined as pairs with $\epsilon< 0$, where $\epsilon$ is the total orbital energy:

\begin{equation}
\epsilon  = \epsilon_{\rm p} + \epsilon_{\rm k} \mbox{,}
\end{equation}

\begin{equation}
\epsilon_{\rm p} = \frac{-G(M_1 + M_2)}{r} \mbox{,} 
\end{equation}

\noindent and

\begin{equation}
\epsilon_{\rm k} = \frac{1}{2}v^2 \mbox{.} 
\end{equation}

\noindent $M_1$ and $M_2$ are the masses of the two sinks, $r$ is the separation between the sinks, and $v$ is their relative velocity.

We find a total of ten binary pairs with orbital separation ranging from a few AU to $\ga$ 1000 AU.  
Some sinks were counted as part of multiple different binaries (i.e. a trinary system), while a total of 15 distinct sinks were in a binary or multiple.  
Given 45 total remaining sinks at this time, we find a binary fraction of 33\%.

\begin{figure}
\includegraphics[width=.45\textwidth]{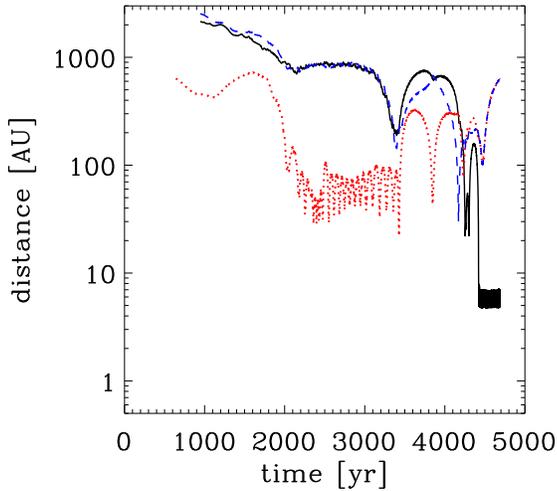}
 \caption{ 
Relative distance between the second and third most massive sink up to the point they form a tight massive binary (solid black line).  
Red dotted line additionally shows the distance between the second-largest sink and the main sink.  
The dashed blue line is the distance between the third-largest sink and the main sink.
 }
\label{sinkdis}
\end{figure}

We additionally note that the second and third-largest sinks of masses 13 and 14 M$_{\odot}$ are orbiting each other with a semi-major axis of $\sim$\,5 AU. 
As shown in Fig. \ref{sinkdis}, the second-largest sink formed a few hundred AU from the main sink.  The third-largest sink formed in the initially more distant secondary clump at $\sim$ 2500 AU from the two largest sinks.  The two largest sinks orbit each other in a $\sim$ 60 AU binary between 2000 and 3000 yr.  At the same time, the sinks from the secondary clump become closer and more intermixed with the sinks from the initial clump, and after 3000 yr the orbit of the two largest sinks becomes disrupted by the resulting N-body dynamics.  
By 4300 yr the second and third largest sinks converge to a tight orbit of 5 AU which continues for the next few hundred years until the end of the simulation.

Future work will further examine how common such close massive binaries were among Pop~III stars (see discussion below). 
In comparison, the lower-resolution and no-feedback simulation of this same minihalo had three binaries with semi-major axes ranging from a few tens to $\sim$\,100 AU. Our increased resolution thus allowed us to form tighter binaries.
The 5 AU separation in our simulation is five times larger than the accretion radius, so angular momentum transport and binary evolution on these scales is well-resolved.  However, the simulation would need to be followed for a longer time to determine whether this binary would later be disrupted by gas inflow or passage of a nearby sink.

\begin{figure*}
\includegraphics[width=.85\textwidth]{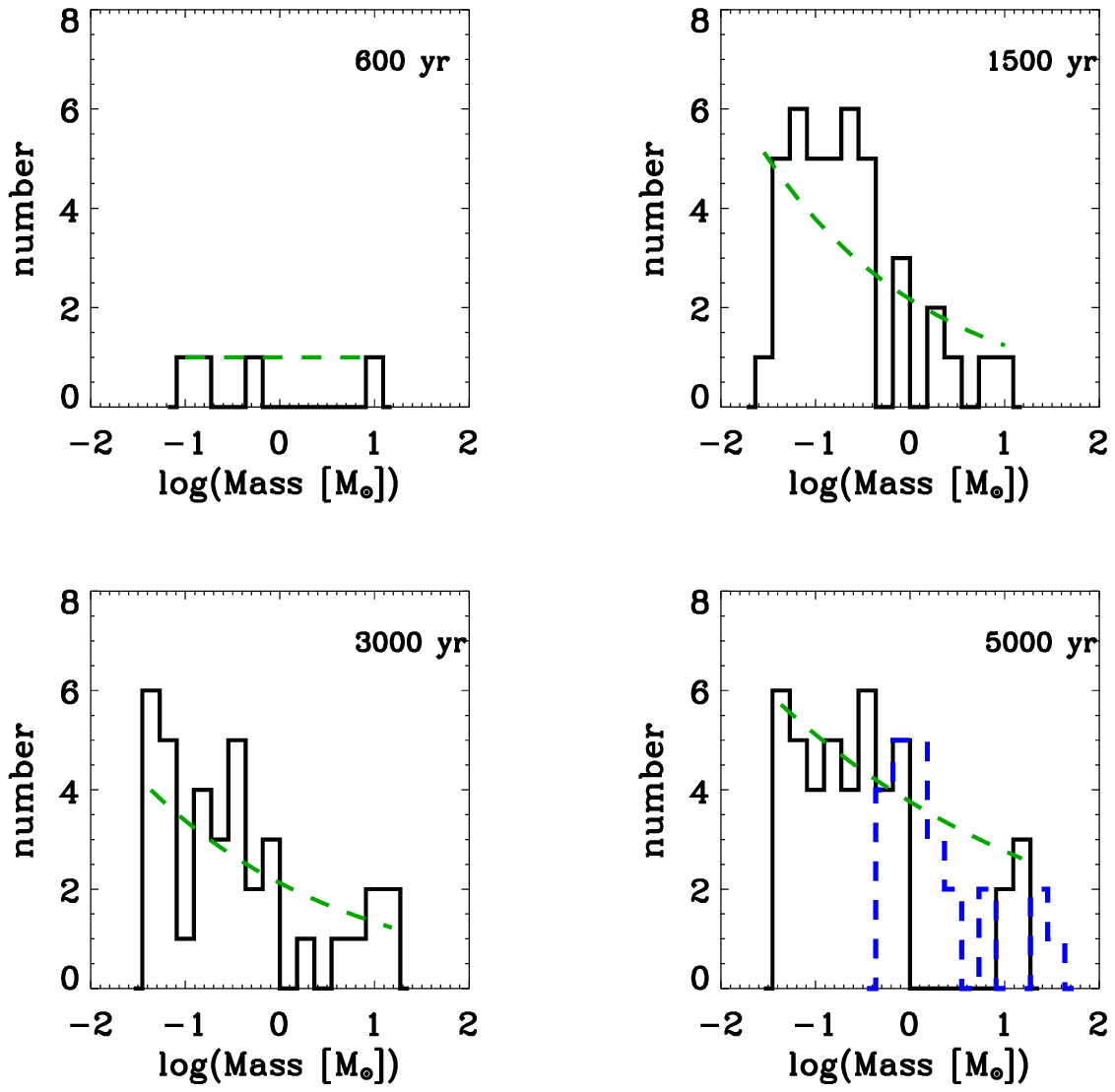}
 \caption{ 
Mass function of sinks shown at four different times throughout the simulation.
Dashed blue line in bottom right panel depicts the mass function of the Pop III cluster when simulated at lower resolution and without feedback (Stacy \& Bromm 2013).
Dashed green lines are least-squares fits to the mass function 
${\rm d}N / {\rm d\,log} \,m \propto  M_* ^ {-x_{\rm MF}}$ found at various times. 
The slopes of these fits are  $x_{\rm MF} =$ 0, 0.24, 0.20, and 0.13 at 600, 1500, 3000, and 5000 yr (i.e.,  $\alpha_{\rm MF} =$ 1, 1.24, 1.20, and 1.13 where ${\rm d}N / {\rm d} m \propto  M_* ^ {-\alpha_{\rm MF}}$) .
}
\label{mhist}
\end{figure*}

\subsection{Mass Function}

The evolution of the mass function of the sinks is shown in Fig. \ref{mhist}.
At early times, before 1000~yr, the small number of sinks can be fit to the function 
\begin{equation}
\frac{ {\rm d}N } { {\rm d}~{\rm log}\,m} \propto  M_* ^ {-x_{\rm MF}} 
\end{equation}
with a flat slope of  $x_{\rm MF}=0$. 
The subsequent rapid formation of large numbers of small sinks, however, leads to a steepening of the mass function.  
A least squares fit gives a slope of $x_{\rm MF} = 0.3$ at 1000~yr, or ${\rm d}N / {\rm d} m \propto  M_* ^ {-\alpha_{\rm MF}}$ where 
$\alpha_{\rm MF}=x_{\rm MF} + 1 = 1.3$.  
At 1500 and 3000~yr, the slope flattens again to $\alpha_{\rm MF}=1.24$ and then to $\alpha_{\rm MF}=1.20$. Towards the end of the simulation, the power-law slope has further decreased to $\alpha_{\rm MF}=1.13$.
This flattening occurs as fewer new low-mass sinks form while the existing sinks continue to accrete mass.  At all times, the mass function is `top-heavy.'  We here define top-heavy as a mass function with $\alpha_{\rm MF} < 2$, such that the majority of protostellar mass will be contained in the most massive stars. For reference, the observed Salpeter slope of the present-day IMF is $\alpha_{\rm MF} = 2.35$.

\cite{susaetal2014} similarly found a top-heavy mass spectrum over several tens of minihaloes, but with a peak at a few tens of solar masses.  
They included radiative feedback but could not resolve the break-out of the ionization front, and they employed a larger sink accretion radius of 30 AU.
As a result, they do not resolve the formation of numerous low-mass stars, and they find no more than six stars forming in a given minihalo.
\cite{hiranoetal2014} additionally include ionization feedback, but cannot follow three-dimensional fragmentation. 
As a result, they follow the growth of a single star in a large number of minihaloes.  This is effectively a distribution function of the most massive star per minihalo, and they find that this maximum mass ranges from 10 to a few hundred solar masses, and is typically on the order of 100 M$_{\odot}$.  \cite{hiranoetal2015} use the primordial cloud characteristics to estimate the maximum Pop III mass for a larger number of minihaloes, finding a distribution with two peaks at $\sim$\,25 and 250 M$_{\odot}$. 
The radiation hydrodynamic simulations of \cite{hosokawaetal2016} are able to follow fragmentation down to scales of 30 AU, and interestingly they find only one star forms in each of their five minihaloes, while 100\% of secondary fragments merge with the central star.  They find the final stellar mass ranges from 15 to several hundred M$_{\odot}$.  
We note here that these previous simulations were followed for longer times of $\sim$ 10$^5$ yr while our simulation was followed for a shorter 5000 yr, which may further contribute to this variation in mass spectra (see also Table 1).

In contrast, our simulation is able to resolve detail at the lower mass end, finding the spectrum extends another order of magnitude down to $\la$ 1 M$_{\odot}$.  We also do not find that the mass spectrum has a central peak at any particular mass.
Furthermore, the $\alpha_{\rm MF}$ values at late times in our simulation are similar to those found in \cite{stacy&bromm2013}.  In particular, \cite{stacy&bromm2013} found that the combined mass function across 10 minihaloes was best fit by $\alpha_{\rm MF} = 0.17$.  In this earlier work, the sinks within each minihalo also accreted for 5000 yr, but protostellar feedback effects were not included.  

Thus, while radiative feedback may alter the final mass reached by the most massive stars, the overall slope of the mass function will not be significantly affected.  In contrast to \cite{stacy&bromm2013} as well as \cite{greifetal2011}, we furthermore find that once rapid fragmentation begins at $\sim$\,1000 yr, the majority of stars have mass $\la$ 1 M$_{\odot}$, and this remains the case for the remainder of the simulation.  
This is not unique to the minihalo we present here.  We point out the 1 AU resolution simulation of a different minihalo presented in \cite{stacy&bromm2014}, in which only five stars formed, all of which grew to only $\sim$\,1 M$_{\odot}$.
 
While some of these small stars within our minihalo may grow further through accretion, this will become increasingly difficult as radiative feedback continues and the I-front expands.  With further accretion or merger events over a longer time, the high-mass end of the mass spectrum may also evolve such that the maximum mass increases to closer to 100 M$_{\odot}$, as seen in previous work.
To summarize, when including three-dimensional feedback physics and using very high numerical resolution, we find a Pop~III mass spectrum that is top-heavy. At the same time, we find a persistent incidence of stars at the low-mass end ($M_*\sim 1$ M$_{\odot}$).

\begin{table*}
\begin{tabular}[width=.95\textwidth]{crrrrrrr}
\hline
Author & d$_{\rm res}$  [AU]   & No. Minialoes & $t_{\rm acc}$ [yr]   &  Feedback   &  $M_{\rm min}$ [M$_{\odot}$]   &   $M_{\rm max}$  [M$_{\odot}$]  &  $M_{\rm med}$ [M$_{\odot}$]   \\
\hline
Stacy ea 2016               &  1.0   &  1          &  5000                  & LW+ion       &   0.05  &   20       &     0.5\\
Stacy \& Bromm 2013   &  20    &  10        &  5000                  & --                &   0.5    &   40        &      2\\
Greif ea 2011                &   0.5   &  5          &  1000                 & Accr. heat   &   0.1    &   10       &       1 \\
Susa ea 2014                &  30    &  59        &  $\sim$10$^5$   & LW             &   0.5     &   200     &      20\\
Hosokawa ea 2016       &  30     &  5        & $\sim$10$^5$    & LW+ion      &   15     &   600      &      300\\
Hirano ea 2014$^{*}$   &  25    &  100     &  $\sim$10$^5$   & LW+ion     &    10     &   2000    &    100\\
\hline 
\end{tabular}
\caption{ Overview of a sample the simulations of primordial star formation discussed in this work.  Columns from left to right are author and year, resolution length,  number of minihaloes followed in the simulation, total protostellar accretion time followed, types of feedback implemented, minimum Pop III mass found in the simulation, maximum mass found, and the approximate median mass. 
We note that the Hirano et al 2014 minihaloes were studied in two dimensions and could not follow three-dimensional fragmentation. 
}
\label{tab1}
\end{table*}

\section{Summary and Conclusions}

We present the first cosmological simulation to indicate that the IMF of the first stars extended to low masses ($\la$ 1 M$_{\odot}$), even under strong feedback effects, while similar to previous studies we find that the IMF will remain top-heavy.  
Our simulation tracked 
the evolution of metal-free gas within a $z \sim 25$ DM minihalo 
We follow the emergence of a Pop III cluster under radiative feedback while resolving scales as small as 1 AU, nearly the maximum size of a Pop III star.
This cluster develops even before an ionization front forms around the most massive star, and it is initially composed of two distinct sub-clusters separated by $\sim$\,2000 AU.  
There are $\sim$ 40 surviving sinks in the disc at the time of I-front breakout, and approximately 30 of these have masses which are $< 1$  M$_{\odot}$.  
A total of $\sim$\,150 sinks formed throughout the primordial disc evolution.
We note that \cite{machida&doi2013} found that in primordial discs the fragmentation radius can be smaller than 1 AU, and that the fragmentation radius has dependence on resolution down to scales of $\sim$ 0.01 AU.  Thus, though we find a relatively large sink formation rate, with increased resolution this number may have been even larger. 
Of the sinks that do form, roughly 1/3 of them survive to the end of the simulation. The rest undergoes mergers with other sinks.  
In addition, 14 of these sinks are ejected from the minihalo, while 11 of the ejected sinks have $M_* < 1$ M$_{\odot}$.

Once the warm phase of neutral gas has expanded to the disc radius of approximately 2000 AU following the LW-induced pressure wave, the accretion rate onto both the disc and sinks slows considerably.  The rate of sink formation and merging declines as well.  However, by this time a high-member cluster with a relatively flat mass function has already formed.  This delay in feedback has important implications for the Pop III mass function and the possibility of observing a truly metal free star in the local Universe.  While a flat mass function is still `top-heavy,' even the most rapidly accreting Pop III clusters can form and eject significant numbers of low-mass stars before an I-front breaks out and mass flow onto the disc is shut off.  

Though feedback has little influence on the protostellar mass function over the initial few thousand years of mass accretion, it will be a determining factor in the maximum mass that can be attained by a given Pop III star (e.g. \citealt{mckee&tan2008, hosokawaetal2011, stacyetal2012, susa2013, hiranoetal2014}).
At the high-mass end of our cluster, we find a massive binary whose members have masses of 13 and 15 M$_{\odot}$ and a separation of 5 AU, while the largest sink within our minihalo reaches a mass of 20 M$_{\odot}$.  However, one or more of these stars may grow closer to 100 M$_{\odot}$ at later times.

The presence of the massive (15 and 13 M$_{\odot}$) binary in our simulation has further implications for Pop III evolution and feedback. 
 If this 5 AU binary grows tighter to the point that a merger or mass transfer would occur, the evolution of both stars would be altered along with their corresponding feedback upon their nearby environment.
If it was common for such high-mass binaries to eventually evolve into a BH binary, then the resulting BH binary would emit X-rays that could ionize more distant gas and influence more distant star formation as well as reionization (e.g. \citealt{jeonetal2012, jeonetal2014}).  We also note the fascinating possibility that Pop III BH binaries will generate detectable gravitational wave signals, and that there is some small probability that such a  signal has already been observed by LIGO.
(e.g. \citealt{aligo2016, kinugawaetal2014, kowalskaetal2012, belczynskietal2004, hartwigetal2016, inayoshietal2016}).

We point out some possible caveats to our results.
For instance, recent work by \nocite{hartwigetal2015b} Hartwig et al. (2015b) introduces a more accurate method to estimate the escape probability of H$_2$ lines in optically thick gas.  Their method implements the TreeCol algorithm.  They find this leads to higher values for escape probability in the density range they consider, up to 
$\sim$\,10$^{15}$ cm$^{-3}$. This promotes lower gas temperature and more fragmentation.  
Using this method would therefore likely strengthen our conclusions.  \cite{greif2014} similarly finds that, compared to a ray-tracing scheme, the Sobolev method underestimates the escape fraction at low optical depths.  
However, at high optical depths 
(roughly when $n > 10^{13}$ cm$^{-3}$), 
this analysis finds that the Sobolev method overestimates the escape fraction.  Nevertheless, due to the strong dependence of the H$_2$ line cooling rate on temperature ($\Lambda_{H_2} \propto T^4$), the different methods lead to minimal change in the thermal evolution of the gas at these densities.  Further studies of the effect of varying methods to calculate the escape probability will be addressed in future work.

Magnetic fields may have a significant effect on the Pop III mass function, as well as the rate of stellar ejections and mergers (e.g. \citealt{machidaetal2006, xuetal2008, machida&doi2013}), particularly if tiny cosmological seed fields can be rapidly amplified within the minihalo through the small-scale dynamo (e.g. \citealt{schleicheretal2010, schoberetal2012}).  Once a protostellar disc is in place, disc dynamos may additionally lead to the development of jets and protostellar outflows (e.g. \citealt{tan&blackman2004, latif&schleicher2015}).  These will be crucial effects to further explore in upcoming simulations.  Such work should also follow the evolution of Pop III clusters for longer than 5000 yr to explore how the mass function changes at later times.  

Along with continually improving simulations, future observations will further constrain the Pop III IMF.
In particular, the low-mass end of the IMF may be revealed through finding Pop III survivors in the Milky Way.  
For instance, \cite{ishiyamaetal2016} use cosmological simulations to conclude that current surveys indicate less than 10 stars per minihalo survive to the present day, while one survivor per minihalo is still consistent with observations. 
Our minihalo produced 11 survivors, above the expected average for all minihaloes.

\nocite{hartwigetal2015a} Hartwig et al (2015a) use semi-analytic models to find that
current surveys may exclude Pop III stars with masses below $M_{\rm min}=$0.65 M$_{\odot}$ at a confidence level of 95\%.  In addition, 
for an IMF that extends to low masses, we should expect to see a 0.8 M$_{\odot}$ Pop III star once surveys sizes of, e.g., Skymapper or Hamburg/ESO reach 10 million stars.  
To compare our results with the predictions of Hartwig et al (2015a), we note that we find a survival fraction of $f_{\rm surv} \sim 0.15$, where $f_{\rm surv}$ is the number of ejected stars with $< 0.8$ M$_{\odot}$ per unit solar mass.
The logarithmically flat IMF used in Hartwig et al. (2015a) instead yields $f_{\rm surv} \sim 0.01$.  Thus increasing their standard prediction by approximately a factor of 10 
(see their fig. 13), we can roughly estimate 10$^6$ Pop III survivors in the Milky Way if our IMF has a minimum mass of 0.1 M$_{\odot}$.
Our results are in some tension with observations, as our simulations indicate low-mass Pop III stars are sufficiently common to have already been observed.
Given our large number of expected survivors, current non-detection may in turn yield $M_{\rm min}$ even above the Hartwig et al (2015a) limit of 0.65 M$_{\odot}$.
It is possible that after ejection, the metal-free nature of Pop III stars may be masked by later accretion of enriched interstellar material (e.g. \citealt{frebeletal2009, johnson&khochfar2011}). 
\cite{johnson2015} finds that some observed carbon-enhanced metal-poor stars may indeed be accretion-polluted Pop III stars.  This may help to make the predictions of Hartwig et al. (2015a) and \cite{ishiyamaetal2016} more consistent with the findings of our work.  
In addition, the high survival fraction derived here may have resulted from
our pre-selection of a vigorously fragmenting minihalo, whereas the
true fraction in an unbiased sample is likely to be lower.

The high-mass end of the IMF may be constrained by observations of PISNe by {\it JWST} (e.g. \citealt{hummeletal2012, panetal2012}), as well as by future facilities, such as the Large Synoptic Survey Telescope (LSST) and the {\it Wide-Field Infrared Survey Telescope (WFIRST)}.  
\cite{hummeletal2012} estimate an upper limit of ~0.2 PISNe per {\it JWST} field of view at any given time, though feedback may reduce this by two orders of magnitude. 
{\it JWST} will also be able to detect CCSNe from lower-mass Pop III progenitors if they occur at $z \sim 10-15$, or even up to $z \sim 20$, if the SN shock collides with a dense circumstellar shell (i.e. Type IIn SNe; Whalen et al. 2013b,a\nocite{whalenetal2013, whalenetal2013b}).

Upcoming observations will additionally test whether the top-heavy mass function we find in our simulation exists in high-$z$ zero-to-low metallicity sources.  
For instance, the COSMOS Redshift 7 (CR7) source is a Ly-$\alpha$ emitter at z=6.6, with Ly $\alpha$ luminosity of $8.3 \times 10^{43}$ erg s$^{-1}$. It has no detected metal emission lines, and it also has a very high HeII line emission of $2 \times 10^{43}$~erg~s$^{-1}$, indicating that the source has a very hard spectrum (e.g. \citealt{sobraletal2015,pallottinietal2015}, \nocite{hartwigetal2015c} Hartwig et al 2015c, \citealt{ smithetal2016}).  
Two guesses for the source of this emission are an accreting black hole (in particular, a direct collapse black hole, DCBH), or a $10^7$ M$_{\odot}$ Pop III stellar cluster.  
We apply our slope of $\alpha_{\rm MF}=1.13$ to such a cluster, assuming an IMF ranging from 0.1 to 200 M$_{\odot}$, 
and we use $L_{\rm zams}$ and $T_{\rm eff}$ values taken from \cite{schaerer2002}.
Normalizing to $10^7$ M$_{\odot}$, we estimate similar Ly$\alpha$ and HeII luminosities of $10^{44}$ and $10^{43}$ erg s$^{-1}$.  The non-ionizing luminosity is of the order $10^{43}$ erg s$^{-1}$, corresponding to measured broadband fluxes of $\la 0.1 \mu$Jy in the $Y$, $J$, and $H$ bands.  
This flux can be easily measured with {\it JWST}, as 10$^4$\,s exposure times allows {\it JWST} to observe a limiting flux of $\sim 10^{-2} \mu$Jy in these bands (see http://www.stsci.edu/jwst/science/sensitivity).  These observations would furthermore improve the current limits on the metallicity of this source. 
{\it JWST} photometry of CR7 as well as other high-$z$ low-metallicity sources promises to greatly improve our picture of stellar cluster and galaxy formation in the early Universe.

Future work will continue to deepen our understanding of the processes that determine the Pop III IMF.  This is crucial to elucidate how Pop III stars shaped early cosmic history. In the coming years we may expect to see theory and simulations work in close concert with observations to shed unprecedented light on the cosmic Dark Ages.

\section*{Acknowledgments}

We thank John Mather, Christopher McKee, and Simon Glover for valuable discussion and feedback.
We are grateful to the referee for many helpful suggestions towards improving this manuscript.
Resources supporting this work were provided by
the NASA High-End Computing (HEC) Program through the
NASA Advanced Supercomputing (NAS) Division at Ames
Research Center.
VB was supported by NSF grant AST-1413501. AS gratefully
acknowledges support through NSF grant AST-1211729 and
by NASA grant NNX13AB84G.
ATL was support by a University of California Dissertation-Year Fellowship.

\bibliographystyle{mn2e}
\bibliography{if_hires}{}

\label{lastpage}

\end{document}